# Operating binary strings using gliders and eaters in reaction-diffusion cellular automaton

Andrew Adamatzky*, Genaro Martinez, Liang Zhang, Andrew Wuensche†

October 25, 2018

Center for Unconventional Computing and Department of Computer Science
University of the West of England, Bristol, United Kingdom

**Abstract**

We study transformations of 2-, 4- and 6-bit numbers in interactions between traveling and stationary localizations in the Spiral Rule reaction-diffusion cellular automaton. The Spiral Rule automaton is a hexagonal ternary-state two-dimensional cellular automaton – a finite-state machine imitation of an activator-inhibitor reaction-diffusion system. The activator is self-inhibited in certain concentrations. The inhibitor dissociates in the absence of the activator. The Spiral Rule cellular automaton has rich spatio-temporal dynamics of traveling (glider) and stationary (eater) patterns. When a glider brushes an eater the eater may slightly change its configuration, which is updated once more every next hit. We encode binary strings in the states of eaters and sequences of gliders. We study what types of binary compositions of binary strings are implementable by sequences of gliders brushing an eater. The models developed will be used in future laboratory designs of reaction-diffusion chemical computers.

## 1 Introduction

Many physical, chemical and biological spatially extended non-linear systems exhibit a wide range of stationary and mobile localizations: solitons, kinks, breathers, excitons, defects and wave-fragments. The localizations can be used to transmit and transform information, and ultimately to perform computation [1]. A unit of information, such as the value of a Boolean variable, is decoded into presence (logical truth) or absence (logical false) of a localization in some specified site of space at a specified moment of time. When two localizations (representing the values of two logical variables) collide, they change their trajectories (or annihilate, reproduce, or change their shape). The new trajectories of the localizations encode the values of some logical function over the two variables. This is how most collision-based computing devices work [7, 10, 13, 1, 8, 19, 20, 14].

*Contact author: andrew.adamatzky@uwe.ac.uk
†www.ddlab.org

The collision-based, or free-space, computing devices typically do not have wires and — in principle – are not supposed to use any other stationary components to perform computation. Any point of the computing media can act as a wire, a trajectory of a traveling localization can be seen as a momentary wire. Any site where two or more localizations collide is a logical gate. Thus space can be used efficiently and nothing is wasted. However, there is a price to pay. Initial positions and launch time of the traveling localizations should be precisely specified: one wrong time step destroys the whole computing scheme. The need for perfect timing is the weakest point of collision-based computing and the subject of constant criticism by "classical" computation schools.

Can we do without perfect timing? Asynchronous cellular-automaton based computers do pretty well [11, 15, 12] by using predetermined wires and valves. In the present paper we are trying to combine pure collision-based computing ideas (gliders only) with stationary architectures (breather-like localization) to implement computing schemes with relaxed timing. We choose the reaction-diffusion Sprial Rule cellular automaton [18, 4, 3, 16] as a testbed for ideas of asynchronous collision-based computing.

What is our rationale behind selecting the Spiral Rule automaton to study novel concepts of the collision-based computing? The Spiral Rule cellular automaton [18] plays a unique role in unconventional computing. On the one hand, this is a simple ternary state hexagonal automaton with Conway's Game of Life type of behavior: it has gliders, still lives and eaters, and glider. Therefore it is very suitable for experimenting with collision-based computing schemes. On the other hand, the Spiral Rule automaton is a unique discrete model of a non-linear reaction-diffusion chemical system with an activator and inhibitor. The gliders and glider guns in the Spiral Rule automaton are analogues of excitation wave-fragments and generators of wave-fragments in a light-sensitive sub-excitable Belousov-Zhabotinsky medium [9]. This means that prototypes of computing schemes designed in the Spiral Rule automaton can then be almost straightforwardly implemented in chemical laboratory prototypes of reaction-diffusion computers.

The paper is structured as follows. The Spiral Rule reaction-diffusion cellular automaton is defined in Sect. 2. Encoding of two- and four-bit binary strings in states of an eater and transformation of the strings by gliders are presented in Sect. 3. Section 4 studies compositions of six-bit binary strings and speculates on the algebraic properties that the compositions impose on a set of six-bit digital numbers. We envisage future developments in Sect. 5.

## 2 Spiral rule reaction-diffusion cellular automaton

A detailed description, including the background and rationale, of the Spiral Rule reaction-diffusion cellular automaton are provided in [18, 4, 3, 16], therefore here we just give a self-consistent outline. The Spiral Rule automaton is a discrete approximation of an abstract reaction-diffusion chemical system. The system has three reactants — activator $A$, inhibitor $I$ and substrate $S$ — and the system is governed by the following set of reactions [18, 4]:

$$
\begin{aligned}
A + 6S &\rightarrow A \qquad & A + I &\rightarrow I \qquad & A + 3I &\rightarrow I \\
& & A + 2I &\rightarrow S & 2A &\rightarrow I \\
& & 3A &\rightarrow A & \beta A &\rightarrow I \\
& & & & I &\rightarrow S.
\end{aligned}
$$

The system is non-linear. The activator is suppressed by the inhibitor when there are small concentrations of the inhibitor and threshold concentrations of the activator. When the inhibitor has a critical concentration (value 2) the inhibitor and activator dissociate producing the substrate.

In the cellular-automaton model each reactant is represented as a cell state. Each cell $x$ of a two-dimensional hexagonal lattice updates its state by the following rule:

$$x^{t+1} = f(\sigma_I(x)^t, \sigma_A(x)^t, \sigma_S(x)^t),$$

where $\sigma_p(x)^t$ is the number of cell $x$'s neighbors with cell-state $p \in \{I, A, S\}$ at time step $t$. Cells update their states synchronously in discrete time-steps. The cell-state transition rule can be compactly represented by matrix $\mathbf{P} = (p_{ij})$, where $0 \leq i \leq j \leq 7$, $0 \leq i + j \leq 7$, $p_{ij} \in \{I, A, S\}$ [3, 18] as follows:

$$x^{t+1} = p_{\sigma_2(x)^t \sigma_1(x)^t}.$$

The matrix has the following structure:

$$
\mathbf{P} = \left\{
\begin{array}{cccccccc}
S & A & I & A & \underline{I} & \underline{I} & \underline{I} & \underline{I} \\
S & I & I & A & \underline{I} & \underline{I} & \underline{I} & \\
S & S & I & A & \underline{I} & \underline{I} & & \\
S & I & I & A & \underline{I} & & & \\
S & S & I & A & & & & \\
S & S & I & & & & & \\
S & S & & & & & & \\
S & & & & & & & \\
 & & & & & & &
\end{array}
\right\}
$$

The cell-state transition matrix could be simplified further because the inhibitor states underlined above are ‘wildcard’ entries [18]. In [4] we demonstrated that each entry of the matrix $\mathbf{P}$ corresponds very well with phenomena in reaction-diffusion systems: diffusion of the activator, suppression of the activator by the inhibitor, self-inhibition of the activator etc.

Any initially random configuration of the Spiral Rule automaton evolves to a dissipative configuration comprised of spiral glider guns (Fig. 1), mobile localizations (gliders) and stationary localized clusters of activator and inhibitor states (eaters). The principle, and of minimal size, localizations are shown in Fig. 2. A glider always has one activator state tailed by a few inhibitor states (Fig. 2, a–c), pretty much as excitation wave-fragment in sub-excitable chemical medium [2]. The eater (Fig. 2d) — in its minimal symmetric form [4] — consists of six activator states surrounded by seven inhibitor states (including the central state of the eater).

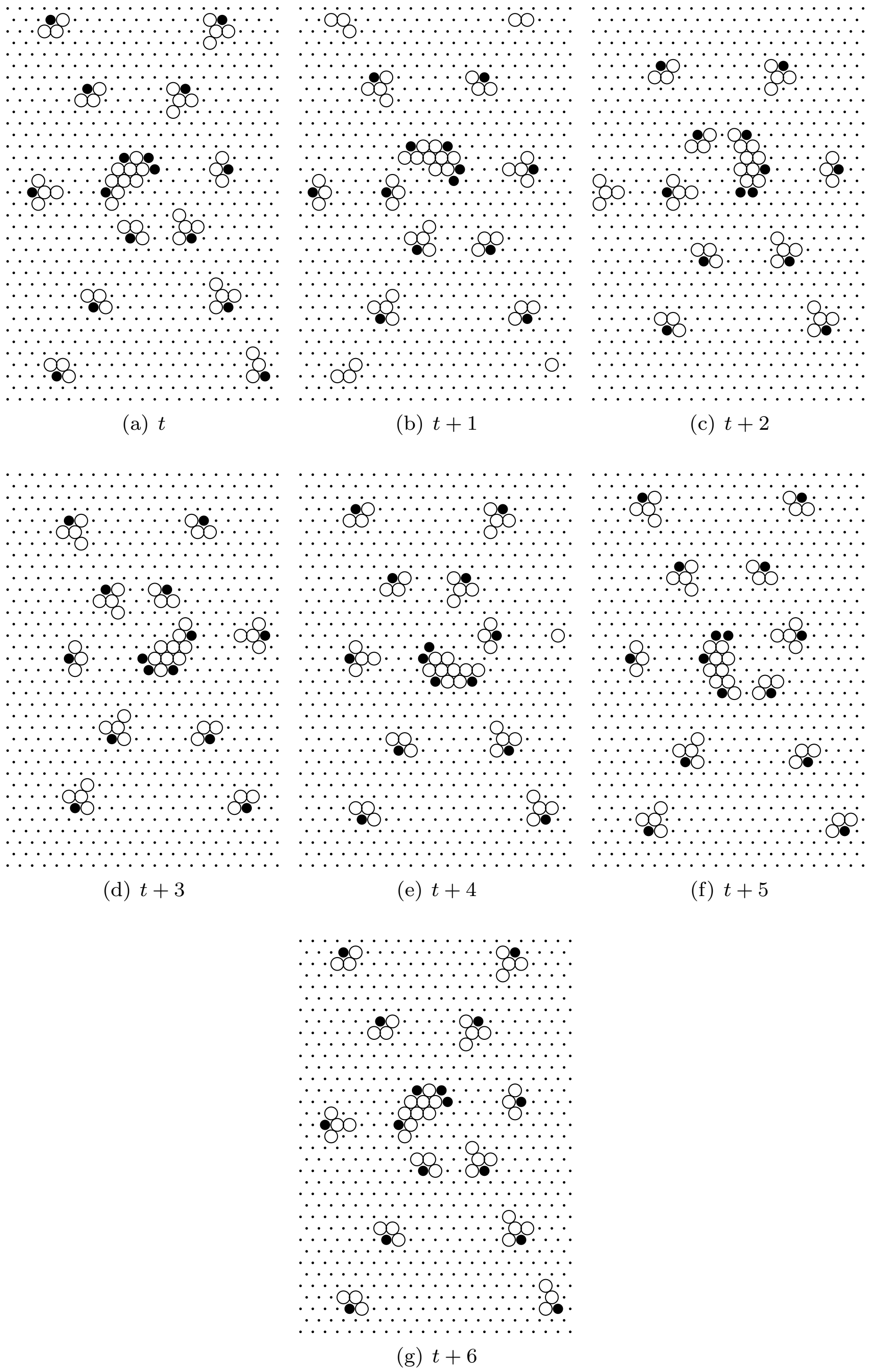

Figure 1: The full cycle of the high-frequency spiral glider gun. The "Core" of the gun is shown rotating (similarly to spiral waves in a sub-excitable Belousov-Zhabotinsky) medium [14]), but can rotate in either direction. The gun emits six streams of gliders. An alternative, low-frequency, spiral glider gun also emerges in the Spiral Rule [18, 16].

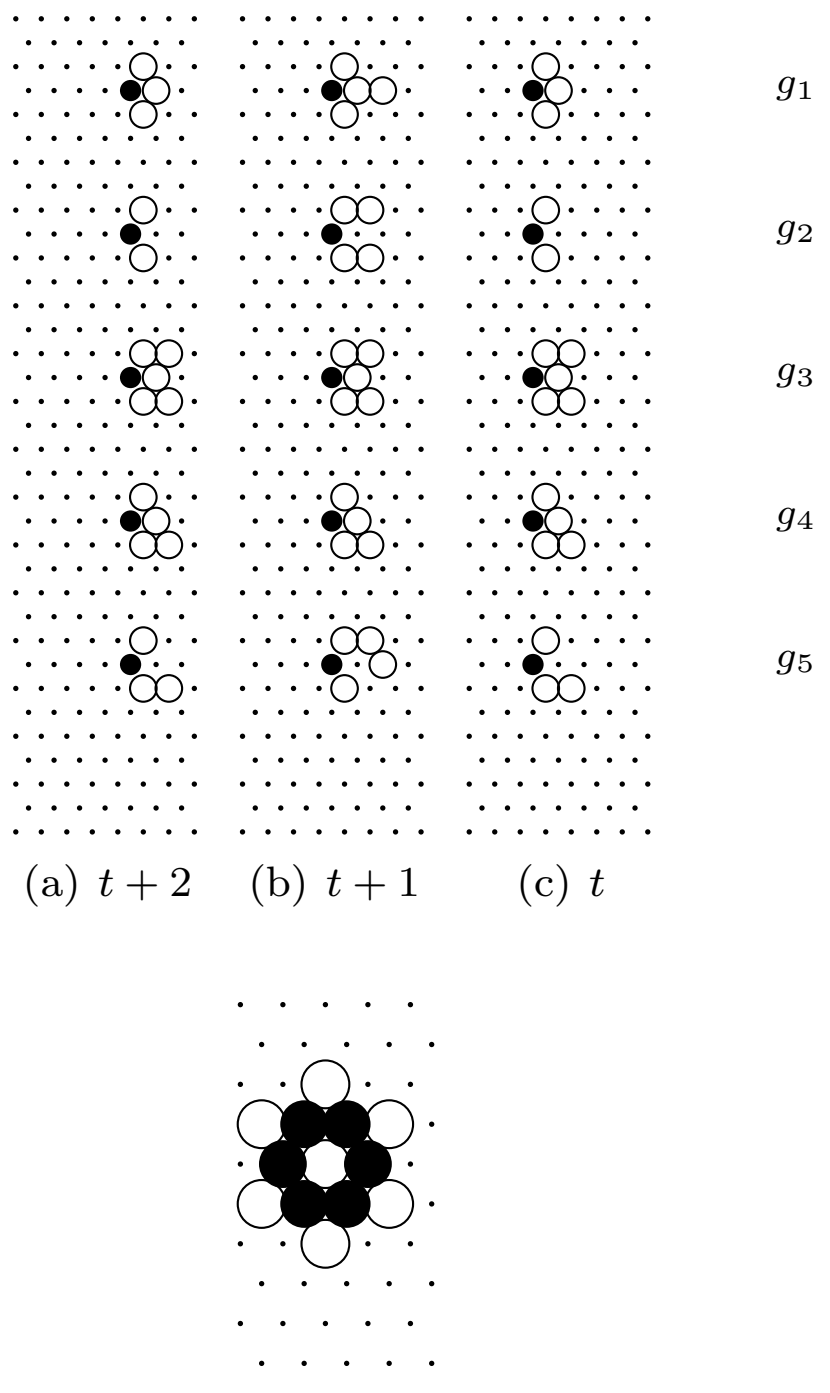

Figure 2: Basic localizations in Spiral Rule. Mobile localizations, gliders types $g1$–$g5$, and the stationary localization, eater (d). Gliders are shown moving from right to left – time steps (c) to (a). Gliders $g1$, $g2$ and $g5$ change shape between alternate time-steps, gliders $g4$ and $g5$ are asymmetric with both chiral forms present. Cell-state $I$ (inhibitor) is shown by a circle, cell-state $A$ (activator) by a solid disk, and cell-state $S$ (substrate) by a dot.

The Spiral Rule supports a rich population of other structures — compound gliders, mobile glider guns, compound eaters [17, 18], in addition to the basic ones discussed. There is also another basic stable localization SL1 [17, 18], which can modify gliders brushing past, but does not have "memory" in its outer shell like the eater discussed in the present paper.

# 3 Transformation of two- and four-bit strings

In [4] we discovered that substrate-sites between inhibitor-sites of the eater can be switched to an inhibitor-state by a colliding glider, or even a glider just brushing past. Thus the eater can play a role of a non-volatile memory device and also implement an asynchronous XOR gate [4]. What if we consider the states of several substrate-sites at once? Then gliders brushing the eater can change these states recursively and thus modify the string content represented by the states.

Let us encode two substrate-states in the northern part of the eater as $x_1$ and $x_2$ (Fig. 3a). If the cell $x_i$ takes the inhibitor state it represents a bit-up state $x_i$ of the string $(x_1x_2)$, if the cell $x_i$ is in the substrate-state then it represents a bit-down state of the string $(x_1x_2)$. We have four possible combination of cell-states $x_1$ and $x_2$: $SS$, $SI$, $IS$, $II$, which corresponds to binary string 00, 01, 10 and 11, respectively (Fig. 3b).

An example of a binary string transformation is shown in Fig. 4. Initially, the eater represents string 00, cells $x_1$ and $x_2$ are in the substrate state (Fig. 4a). We launch glider $g_1$ traveling West (Fig. 4b). When the glider brushes along the outer edge of the eater (Fig. 4a)f–k) it causes a minor perturbation of the marginal cell-states. The result of the perturbation is that one substrate-cell $x_1$ becomes an inhibitor (Fig. 4a)k). The glider changes its state as well — $g_1 \rightarrow g_5$ — because of its interaction with the eater (Fig. 4a)j–o).

All configurations, including those modified by gliders, are still configurations, they are not evolving by themselves but can only be changed again by another brushing glider. When we apply a sequence of transformation we do not have to worry about the precise timing of glider generation or a time interval between gliders (as far as the gliderphase is preserved). Therefore, eater-glider based transformations can be classified as asynchronous.

**Finding 1** *The following transformations of two-bit strings, encoded below in digital numbers, are implemented by gliders and eaters in the Spiral Rule cellular automaton:*

| | | |
|---|---|---|
| $L_1$ : | $L_2$ : | |
| $0 \rightarrow 0$ | $0 \rightarrow 2 \rightarrow 0$ | $L_3 : 0 \rightarrow 1 \rightarrow 2 \rightarrow 3 \rightarrow 0$ |
| $1 \rightarrow 3 \rightarrow 1$ | $1 \rightarrow 1$ | $L_4 : 0 \rightarrow 3 \rightarrow 2 \rightarrow 1 \rightarrow 0$ |
| $2 \rightarrow 2$ | $3 \rightarrow 3$ | |

Assuming states '0' and '1' are Boolean values, transformations $L_1 \cdots L_3$ can be expressed in the following manner:

$$L_1(x_i, x_{i-1}) = (x_i \vee y_i \wedge (x_i \oplus x_{i-1}), x_{i-1} \vee y_i \wedge x_{i-1}) \quad (1)$$

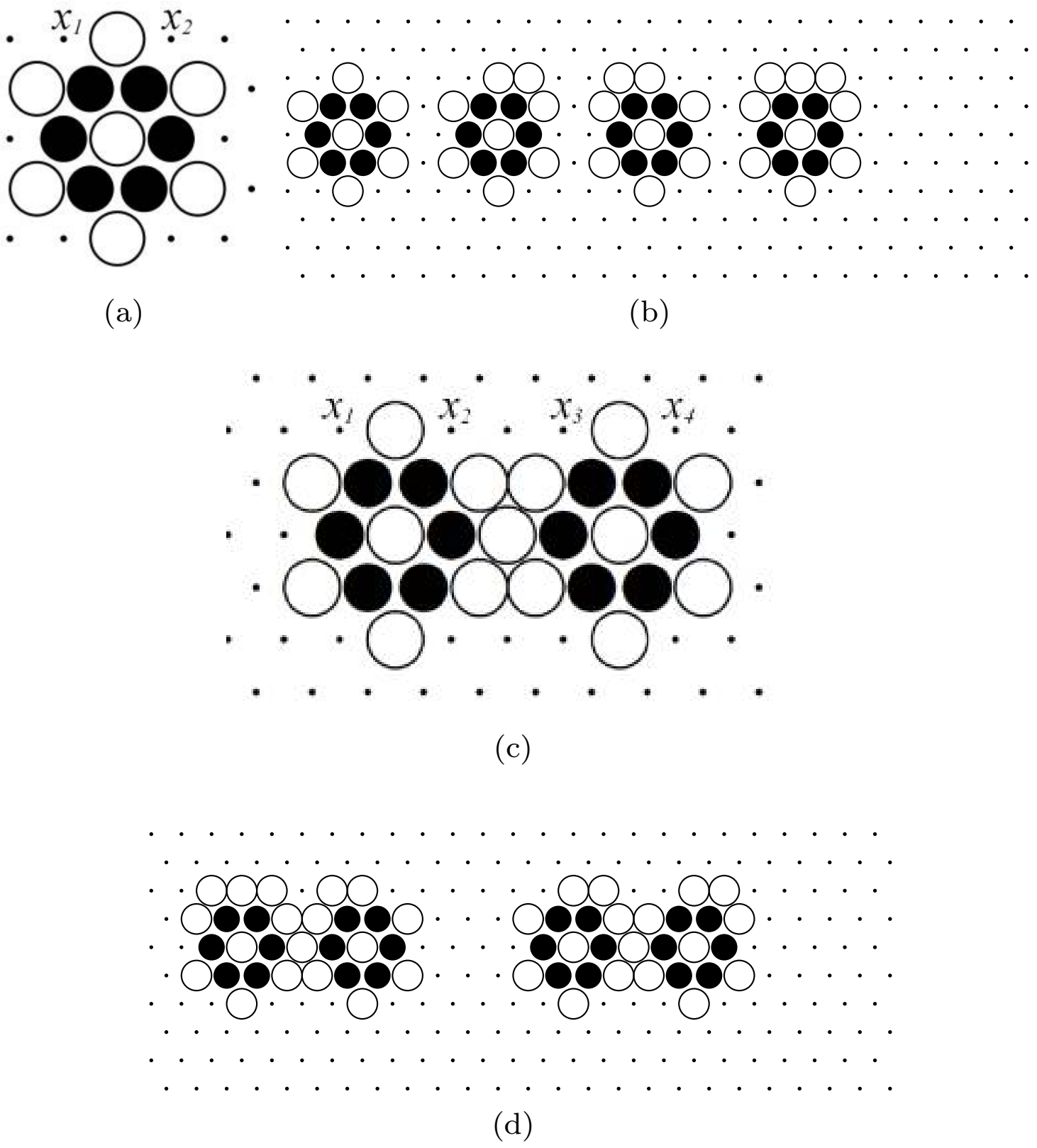

Figure 3: Encoding two-bit strings in the states of eaters. The state of cells $x_1$ and $x_4$ for a single eater encoding (a) and cells $x_1 \dots x_4$ for the eater couple (c) encoding represent bits of strings $(x_1 x_2)$ and $(x_1 x_2 x_3 x_4)$ respectively. Cell-state $I$ (inhibitor) is shown by a circle, cell-state $A$ (activator) by a solid disk, and cell-state $S$ (substrate) by a dot. Cell $x_i$ in state $I$ represents '1', cell in state $S$ represents '0'. Examples of encoding strings 00, 01, 10, 11 in a single eater are shown in (b). Examples of encoding strings 1110 and 0110 in the eater couple are shown in (d).

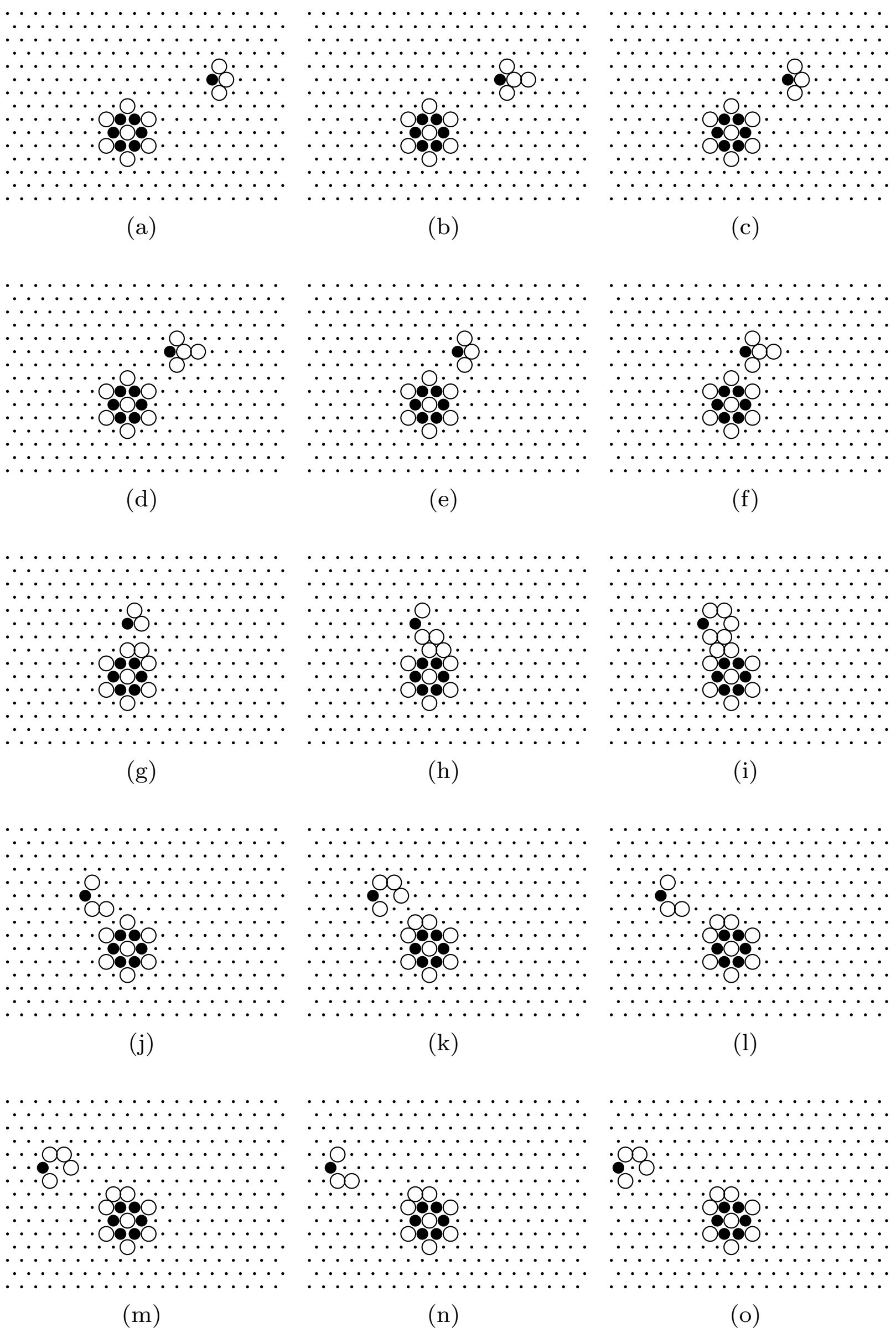

Figure 4: Transformation $00 \to 10$ implemented by glider $g_1$.

$$L_2(x_i, x_{i-1}) = (x_i \vee y_i \wedge (\overline{x_i \oplus x_{i-1}}), x_{i-1} \vee y_i \wedge x_{i-1}) \quad (2)$$

$$L_3(x_i, x_{i-1}) = (x_i \vee y_i \wedge (x_i \oplus x_{i-1}), x_{i-1} \vee y_i \wedge \overline{x_{i-1}}) \quad (3)$$

$$L_4(x_i, x_{i-1}) = (x_i \vee y_i \wedge (\overline{x_i \oplus x_{i-1}}), x_{i-1} \vee y_i \wedge \overline{x_{i-1}}). \quad (4)$$

**Finding 2** *One can implement increment modulo 4 using transformation $L_3$ and decrement modulo 4 using transformation $L_3$*

This follows directly from Finding 1 and Boolean representation (3).

Similarly by positioning two eaters close to each other — the distance between central states of eaters is 4 — we can change four substrate-states just with one glider. The positions of substrate-states $x_1 \dots x_4$ used for encoding are shown in Fig. 3c; examples of encoding four-bit binary sequences are demonstrated in Fig. 3d.

Example of transformation 0000 → 1100 is shown in Fig. 5, and transformation 1100 → 0010 in Fig. 6.

**Finding 3** *The following transformations of four-bit strings, encoded below in digital numbers, are implemented by gliders and eaters in the Spiral Rule cellular automaton:*

| $T_1$ : | $T_2$ : |
|---|---|
| $0 \to 12 \to 8 \to 4 \to 0$ | $0 \to 6 \to 4 \to 10 \to 0$ |
| $1 \to 3 \to 13 \to 7 \to 1$ | $1 \to 9 \to 1$ |
| $2 \to 2$ | $2 \to 8 \to 14 \to 12 \to 2$ |
| $5 \to 15 \to 9 \to 11 \to 5$ | $3 \to 7 \to 11 \to 15 \to 3$ |
| $6 \to 14 \to 6$ | $5 \to 5$ |
| $10 \to 10$ | $13 \to 13$ |

$T3$ :
$0 \to 9 \to 14 \to 3 \to 8 \to 1 \to 6 \to 11 \to 0$
$2 \to 7 \to 4 \to 5 \to 10 \to 15 \to 12 \to 13 \to 2$

$T4$ :
$0 \to 3 \to 2 \to 13 \to 8 \to 11 \to 10 \to 5 \to 0$
$1 \to 12 \to 7 \to 14 \to 9 \to 4 \to 15 \to 6 \to 1$

The exact conditions $C(T_i)$ or $C(L_i)$ of transformations $T_i$ and $L_i$ are determined by two parameters $(g, d)$, where a type $g$ glider brushes the eater, and the initial distance $d$ between the eater and the glider is even.

**Finding 4** *The conditions of the transformations $L_i$ and $T_i$, $i = 1 \dots 4$ are as follows:*

- $C(L_1) = C(T_1) = (g', d')$, $g' = g_3, g_4$ *and* $d'$=*even, odd.*

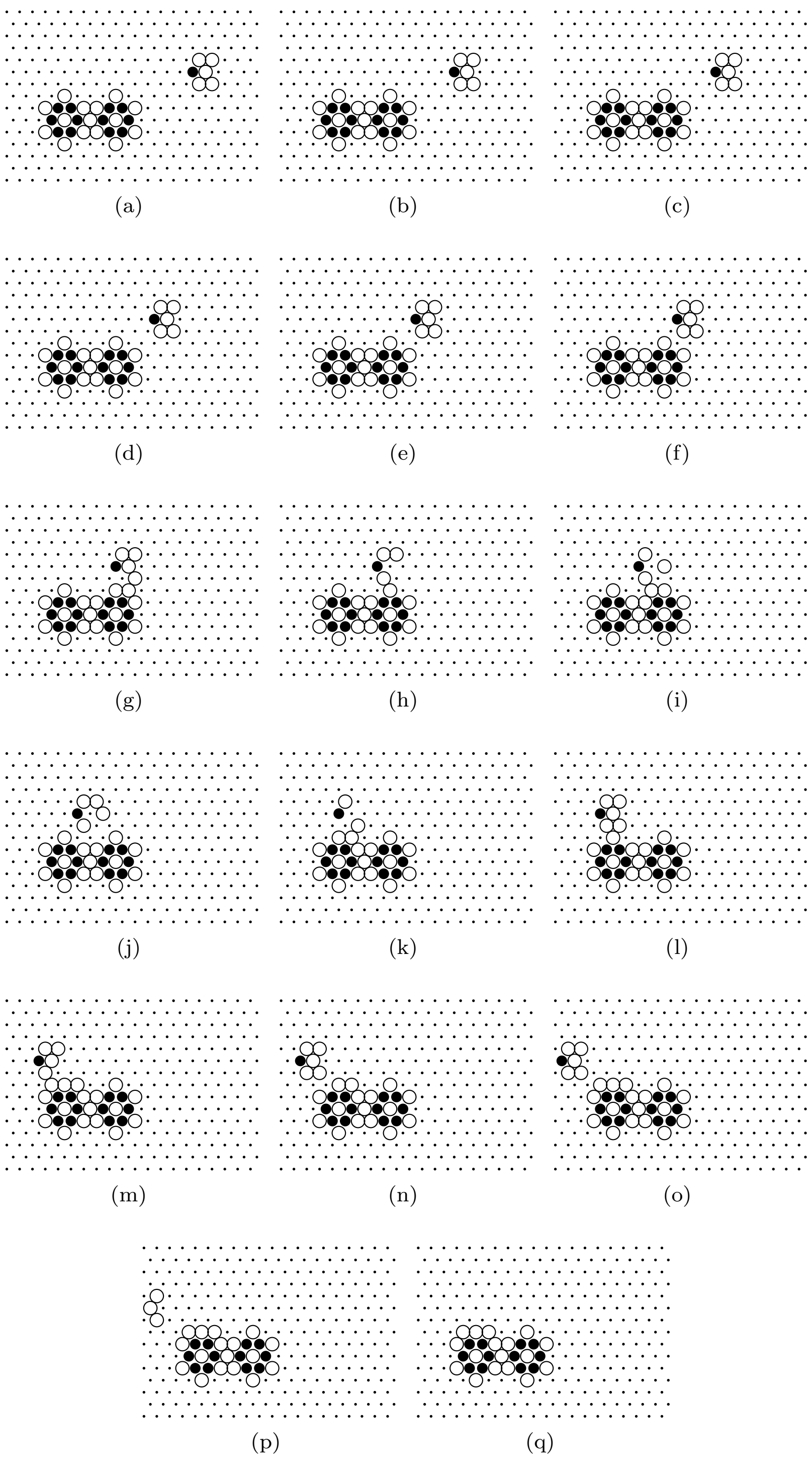

(a) (b) (c) (d) (e) (f) (g) (h) (i) (j) (k) (l) (m) (n) (o) (p) (q)

Figure 5: Transformation 0000 → 1100 implemented by glider $g_3$.

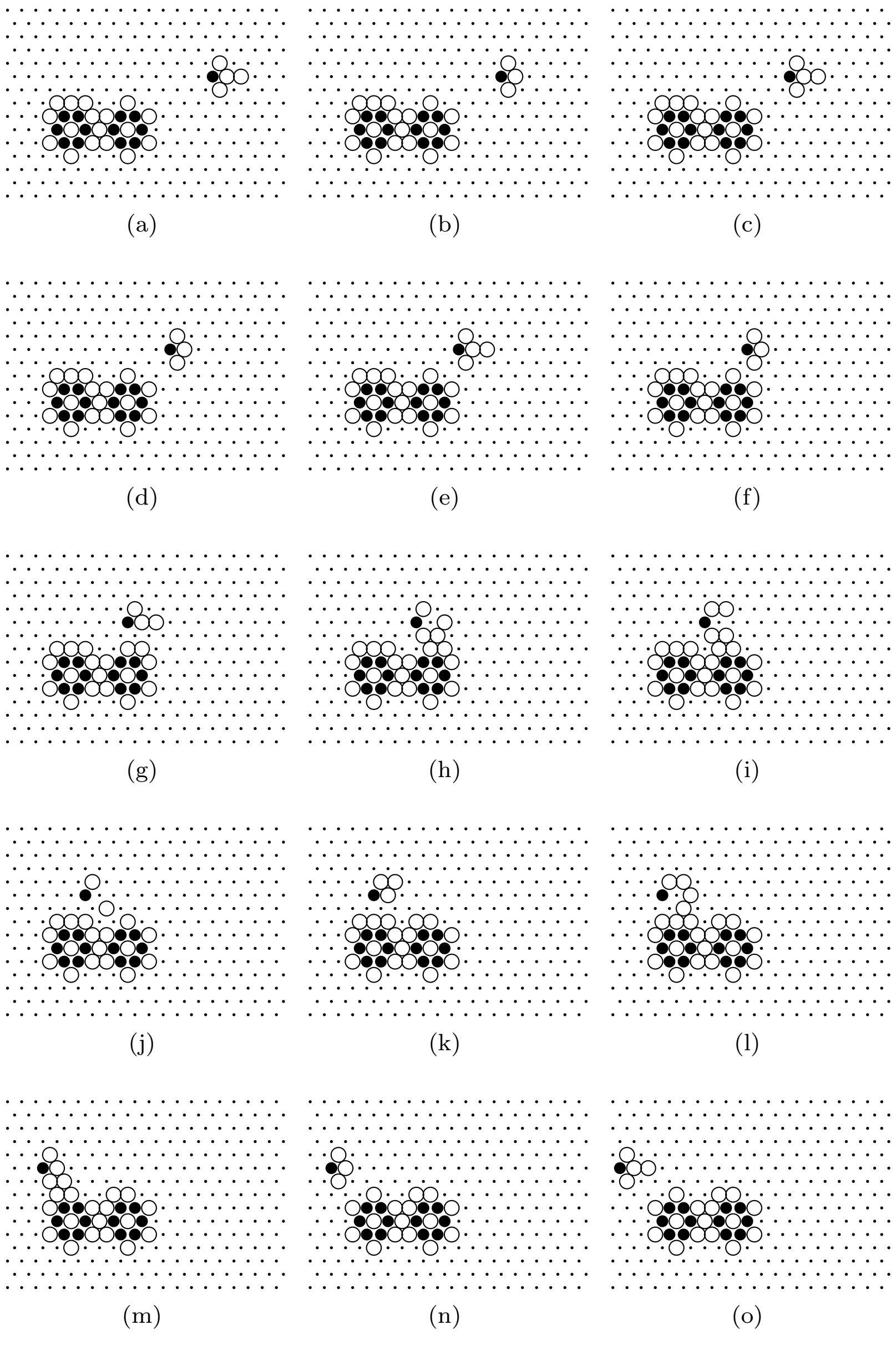

Figure 6: Transformation 1100 → 0010 implemented by glider $g_1$.

- $C(L_2) = C(T_2) = (g', d')$, $g' = g_1$ *and* $d'$=*even, odd.*
- $C(L_3) = C(T_3) = \{(g_2, \textit{even}), (g_7, \textit{even}), (g_5, \textit{odd})\}$
- $C(L_4) = C(T_4) = \{(g_2, \textit{odd}), (g_7, \textit{odd}), (g_5, \textit{even})\}$.

**Finding 5** *Transformations* $T_1$ *and* $T_2$ *split the set* $\{0, \cdots, 15\}$ *into stable subsets of exclusively even numbers and subsets of exclusively odd numbers.*

This is an outcome of the experimental implementation of the transformations $T_1$ and $T_2$, and analysis of chains generated by the transformations in Finding 3.

**Finding 6** *By using a series of transformations* $T_1, \ldots, T_4$ *one can transform a number from* $\{0, \ldots, 15\}$ *to any other number from* $\{0, \ldots, 15\}$ *in at most three steps.*

To prove this statement we compute a distance matrix $D = (d_{ij})_{0 \leq i,j \leq 15}$, where $d_{ij}$ represents the minimal number of transformations $T_1, \ldots T_4$ that one needs to apply to get from number $i$ to number $j$. The distance matrix has the following form:

$$D = \begin{pmatrix}
3&2&2&1&2&3&1&2&2&1&3&2&1&2&2&3\\
3&2&2&1&2&3&1&2&2&1&3&2&1&2&2&3\\
3&2&1&3&2&3&3&1&1&3&3&2&2&1&2&3\\
3&2&1&3&2&3&3&1&1&3&3&2&2&1&2&3\\
1&3&3&2&3&1&2&3&3&2&1&3&2&3&3&1\\
1&3&3&2&3&1&2&3&3&2&1&3&2&3&3&1\\
2&1&3&2&1&2&2&3&3&2&2&1&2&3&1&2\\
2&1&3&2&1&2&2&3&3&2&2&1&2&3&1&2\\
2&1&3&2&1&2&2&3&3&2&2&1&2&3&1&2\\
2&1&3&2&1&2&2&3&3&2&2&1&2&3&1&2\\
1&3&3&2&3&1&2&3&3&2&1&3&2&3&3&1\\
1&3&3&2&3&1&2&3&3&2&1&3&2&3&3&1\\
3&2&1&3&2&3&3&1&1&3&3&2&2&1&2&3\\
3&2&1&3&2&3&3&1&1&3&3&2&1&2&2&3\\
3&2&2&1&2&3&1&2&2&1&3&2&1&2&2&3\\
3&2&2&1&2&3&1&2&2&1&3&2&1&2&2&3
\end{pmatrix}$$

For example, to get 8 from 7 we must apply $T_1$ twice and then $T_3$ once: $T_3(T_1(T_1(7))) = 8$. From $D$ we can also tell that one needs 34 to generate a glider, which brushes the eater to generate a monotonous ordered sequence of numbers $(0, 1, \ldots, 15)$.

# 4 Six bit coding

Let us now consider the composition $\mathbf{x} \circ \mathbf{y}$ of two six bit binary strings $\mathbf{x} = (x_1 \cdot x_6)$ and $\mathbf{y} = (y_1 \cdot y_6)$. The string $\mathbf{x}$ is represented by states of six cells $x_1 \cdots x_6$ surrounding the eater, as shown in Fig. 7a. The string $\mathbf{y}$ is represented by gliders brushing the eater (as discussed in the previous sections). The trajectories of gliders are indicated in Fig. 7: $y_i = 1$ if there is a glider traveling along trajectory $y_i$ (Fig. 7)a. Examples of encoding binary strings in the eater's states are shown in (Fig. 7)b.

Gliders representing string $\mathbf{y}$ are sent in the order $y_1, \cdots, y_6$. Values of Boolean variables $y_i$, $i = 1, \cdots, 6$, are represented by the presence or absence of a glider $g$, traveling along trajectory $y_i$. A minimum interval between two subsequent gliders is selected so the next glider is sent after the current glider has already passed the eater. There is no maximal interval between the gliders. The exact structure of composition $\mathbf{x} \circ \mathbf{y}$ is determined by transformations $L_1 \cdots L_4$.

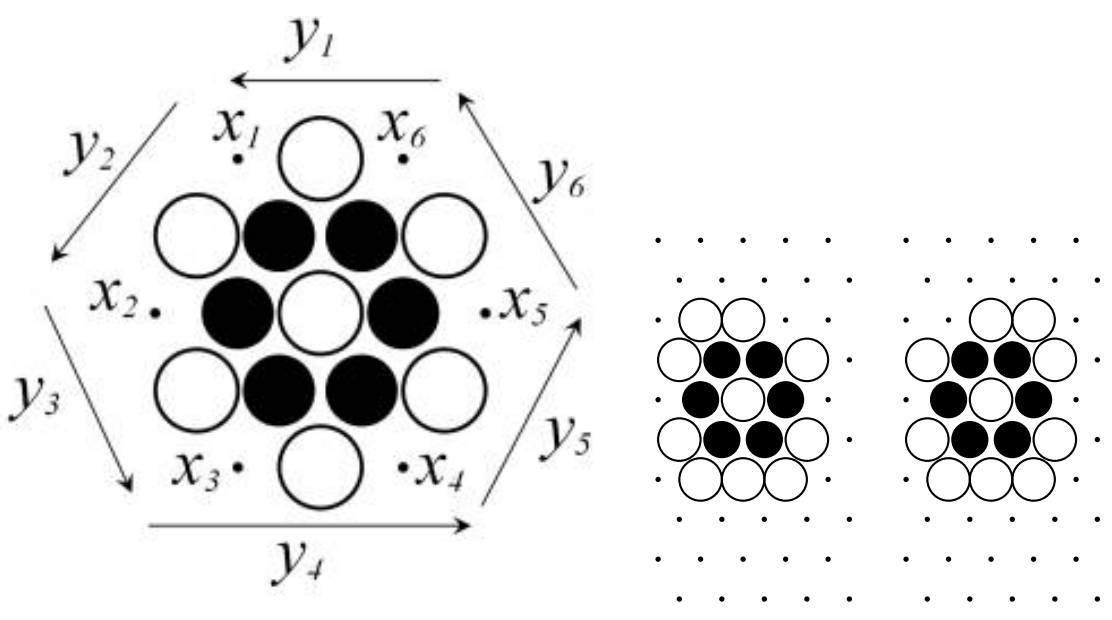

(a) (b) (c)

Figure 7: Encoding six-bit strings in states of the eaters. (a) States of cells $x_1 \cdots x_6$ represent bits of strings $(x_1 \cdots x_6)$. Cell-state $I$ (inhibitor) is shown by a black disk, cell-state $A$ (activator) by a circle, and cell-state $S$ (substrate) by a dot. Cell $x_i$ in state $I$ represents '1', cell in state $S$ represents '0'. (b)–(c) Examples of encoding strings 101100 (a) and 001101 (b). Vectors $y_1 \cdots y_6$ in (a) show which direction gliders representing input string $y_1 \cdots y_6$ travel.

For example, if the composition $\mathbf{x} \circ \mathbf{y}$ is based on $L_4$ then the result $\mathbf{x}' \leftarrow \mathbf{x} \circ \mathbf{y}$ can be calculated as follows:

$$x_1' = x_1 \vee y_1 \wedge (\overline{x_1 \oplus x_6})$$
$$x_6' = x_6 \vee y_1 \wedge \overline{x_6}$$

and for $i = 2, \dots, 6$

$$x_i' = x_i \vee y_i \wedge (\overline{x_i \oplus x_{i-1}})$$
$$x_i' = x_{i-1} \vee y_i \wedge \overline{x_{i-1}}$$

We represent the composition $\mathbf{x} \circ \mathbf{y}$ implemented by the eater and glider $g_z$, $z = 1, \cdots, 6$, by matrix $M(L_i) = (m_{ij})_{0 \leq i,j \leq 63}$. Each entry $m_{ij}$ of the matrix is an integer from $\{0, \cdots, 63\}$. The entry $m_{ij}$ is a digital representation of the string $\mathbf{x} \leftarrow \mathbf{x} \circ \mathbf{y}$, and its indexes $i$ and $j$ are digital representations of strings $\mathbf{x}$ and $\mathbf{y}$, respectively.

Matrices of the compositions $\mathbf{x} \circ \mathbf{y}$ are shown in Appendix (Fig. 8–11). Structures of the matrices do not indicate any immediate implications for practical applications. There is however a range of properties which may be useful in further implementations of arithmetical operations with gliders and eaters.

**Finding 7** *Let $\kappa(L_i)$, $i = 1, \dots, 4$, be a ratio of commutative pairs in $\mathbf{x} \circ \mathbf{y}$ determined by transformation $L_i$, and $\alpha(L_i)$ a ratio of associative triples, then the following hierarchy of compositions can be drawn:*

$$\kappa(L_3) > \kappa(L_1) > \kappa(L_4) > \kappa(L_2)$$

$$\alpha(L_2) > \alpha(L_3) > \alpha(L_1) > \alpha(L_4)$$

This follows from straightforward analysis of the matrices $M(L_1), \dots, M(L_4)$ in Tab. 1.

**Finding 8** *For compositions determined by transformations $L_1$ and $L_2$ we have* $\mathbf{x} \circ \mathbf{x} = \mathbf{x}$ *for any* $\mathbf{x}$ *from* $\{0, \dots, 63\}$.

| $L_i$ | $\kappa(L_i)$ | $\alpha(L_i)$ |
|---|---|---|
| $L_1$ | 0.33 | 0.79 |
| $L_2$ | 0.11 | 0.85 |
| $L_3$ | 0.35 | 0.81 |
| $L_4$ | 0.15 | 0.63 |

Table 1: Ratios of commutative pairs $\kappa(\cdot)$ and associative triples $\alpha(\cdot)$ in $M(L_1), \ldots, M(L_4)$.

The following finding relates to identities and absorbing elements (nulls) of the compositions.

**Finding 9** *Number 63 is the absorbing element of compositions determined by $L_3$ and $L_4$. The composition determined by $L_1$ has two left absorbing elements: 0 and 63. The composition determined by $L_2$ has eighteen left absorbing elements: 21, 23, 27, 29, 31, 42, 43, 45, 46, 47, 53, 54, 55, 58, 59, 61, 62, 63. None of the composition has left or right identities.*

The structure of the first two rows of $M(L_1)$ shows that stationary and traveling localizations in the Spiral Rule automaton can implement at least one sensible arithmetical operation: transformation of arbitrary stream of numbers to streams of even or odd numbers.

**Finding 10** *Let number $a \in \{0, 1\}$, encoded in a state of the eater, and number $b \in \{0, \ldots, 31\}$, encoded by a stream of gliders traveling along trajectories as as per Fig. 7, then the eater-gliders system executes operation $2b + a$; the result of the operation is encoded in state of the eater.*

The resultant state of the eater can be read and erased by additional stream of gliders, as have been already demonstrated in [4].

# 5 Discussion

Using a cellular-automaton model [4] of a reaction-diffusion system we uncovered ways of implementing asynchronous operations over sets of binary strings. We employ a sequence of mobile localizations (gliders, excitation wave-fragments) to represent one binary string, and a stationary localization (eater, breather, standing wave) to represent another string in the composition. The operations are asynchronous in the sense that there is no precise time for a glider to collide with the eater, and there is no fixed time interval between gliders in a glider train representing a binary string. The order in which gliders hit certain parts of the eater is nevertheless important and must be executed precisely during the computation.

We found that by colliding gliders with eaters we can implement transformations of sets of two-, four- and six-bit integers, which demonstrates collision based computation in the Spiral Rule. The methods could find applications for the design arithmetical chips.

We did not discover any sophisticated operations on digital numbers. Most impressive so far were increments and decrements modulo four and subdivision of the digital number set onto prime and even numbers. We found operations on a set of six-bit integers which are rich yet irregular and require detailed study. Nevertheless are we convinced that by combining operations (based on $L_1, \ldots, L_4$) one can implement typical arithmetical operations. This will be subject of further analysis.

What are the potential practical outcomes of our designs and findings? As we demonstrated in [14], the Spiral Rule automaton is an ideal discrete approximation of excitation dynamics in a light-sensitive sub-excitable chemical medium on a patterned environment. Gliders are discrete analogs of localized wave-fragments. This does not mean that our computing schemes can be implemented directly in an excitable chemical medium, not least because there are no excitation-wave analogues of eaters. More feasible implementations would be in other substrates, possibly at the nano-scales as e.g. excitation patterns in networks of single-electron oscillators [5], excitons in mono-molecular layers and breathers in polymers and crystals [1], and dynamically evolving molecular circuits [6].

## Appendix

| | | | | | | | | | | | | | | | | | | | | | | | | | | | | | | | | | | | | | | | | | | | | | | | | | | | | | | | | | | | | | | | |
|---|---|---|---|---|---|---|---|---|---|---|---|---|---|---|---|---|---|---|---|---|---|---|---|---|---|---|---|---|---|---|---|---|---|---|---|---|---|---|---|---|---|---|---|---|---|---|---|---|---|---|---|---|---|---|---|---|---|---|---|---|---|---|---|
|0|0|0|0|0|0|0|0|0|0|0|0|0|0|0|0|0|0|0|0|0|0|0|0|0|0|0|0|0|0|0|0|0|0|0|0|0|0|0|0|0|0|0|0|0|0|0|0|0|0|0|0|0|0|0|0|0|0|0|0|0|0|0|0|
|1|1|1|1|1|1|1|1|1|1|1|1|1|1|1|1|1|1|1|1|1|1|1|1|1|1|1|1|1|1|1|1|33|33|33|33|33|33|33|33|33|33|33|33|33|33|33|33|49|49|49|49|49|49|49|49|57|57|57|57|61|61|63|63|
|2|3|2|3|2|3|2|3|2|3|2|3|2|3|2|3|2|3|2|3|2|3|2|3|2|3|2|3|2|3|2|3|2|3|2|3|2|3|2|3|2|3|2|3|2|3|2|3|2|3|2|3|2|3|2|3|2|3|2|3|2|3|2|3|
|3|3|3|3|3|3|3|3|3|3|3|3|3|3|3|3|3|3|3|3|3|3|3|3|3|3|3|3|3|3|3|3|35|35|35|35|35|35|35|35|35|35|35|35|35|35|35|35|51|51|51|51|51|51|51|51|59|59|59|59|63|63|63|63|
|4|4|6|7|4|4|6|7|4|4|6|7|4|4|6|7|4|4|6|7|4|4|6|7|4|4|6|7|4|4|6|7|4|4|6|7|4|4|6|7|4|4|6|7|4|4|6|7|4|4|6|7|4|4|6|7|4|4|6|7|4|4|6|7|
|5|5|7|7|5|5|7|7|5|5|7|7|5|5|7|7|5|5|7|7|5|5|7|7|5|5|7|7|5|5|7|7|37|37|39|39|37|37|39|39|37|37|39|39|37|37|39|39|53|53|55|55|53|53|55|55|61|61|63|63|61|61|63|63|
|6|7|6|7|6|7|6|7|6|7|6|7|6|7|6|7|6|7|6|7|6|7|6|7|6|7|6|7|6|7|6|7|6|7|6|7|6|7|6|7|6|7|6|7|6|7|6|7|6|7|6|7|6|7|6|7|6|7|6|7|6|7|6|7|
|7|7|7|7|7|7|7|7|7|7|7|7|7|7|7|7|7|7|7|7|7|7|7|7|7|7|7|7|7|7|7|7|39|39|39|39|39|39|39|39|39|39|39|39|39|39|39|39|55|55|55|55|55|55|55|55|63|63|63|63|63|63|63|63|
|8|8|8|8|12|12|14|15|8|8|8|8|12|12|14|15|8|8|8|8|12|12|14|15|8|8|8|8|12|12|14|15|8|8|8|8|12|12|14|15|8|8|8|8|12|12|14|15|8|8|8|8|12|12|14|15|8|8|8|8|12|12|14|15|
|9|9|9|9|13|13|15|15|9|9|9|9|13|13|15|15|9|9|9|9|13|13|15|15|9|9|9|9|13|13|15|15|41|41|41|41|45|45|47|47|41|41|41|41|45|45|47|47|57|57|57|57|61|61|63|63|57|57|57|57|61|61|63|63|
|10|11|10|11|14|15|14|15|10|11|10|11|14|15|14|15|10|11|10|11|14|15|14|15|10|11|10|11|14|15|14|15|10|11|10|11|14|15|14|15|10|11|10|11|14|15|14|15|10|11|10|11|14|15|14|15|10|11|10|11|14|15|14|15|
|11|11|11|11|15|15|15|15|11|11|11|11|15|15|15|15|11|11|11|11|15|15|15|15|11|11|11|11|15|15|15|15|43|43|43|43|47|47|47|47|43|43|43|43|47|47|47|47|59|59|59|59|63|63|63|63|59|59|59|59|63|63|63|63|
|12|12|14|15|12|12|14|15|12|12|14|15|12|12|14|15|12|12|14|15|12|12|14|15|12|12|14|15|12|12|14|15|12|12|14|15|12|12|14|15|12|12|14|15|12|12|14|15|12|12|14|15|12|12|14|15|12|12|14|15|12|12|14|15|
|13|13|15|15|13|13|15|15|13|13|15|15|13|13|15|15|13|13|15|15|13|13|15|15|13|13|15|15|13|13|15|15|45|45|47|47|45|45|47|47|45|45|47|47|45|45|47|47|61|61|63|63|61|61|63|63|61|61|63|63|61|61|63|63|
|14|15|14|15|14|15|14|15|14|15|14|15|14|15|14|15|14|15|14|15|14|15|14|15|14|15|14|15|14|15|14|15|14|15|14|15|14|15|14|15|14|15|14|15|14|15|14|15|14|15|14|15|14|15|14|15|14|15|14|15|14|15|14|15|
|15|15|15|15|15|15|15|15|15|15|15|15|15|15|15|15|15|15|15|15|15|15|15|15|15|15|15|15|15|15|15|15|47|47|47|47|47|47|47|47|47|47|47|47|47|47|47|47|63|63|63|63|63|63|63|63|63|63|63|63|63|63|63|63|
|16|16|16|16|16|16|16|16|24|24|24|24|28|28|30|31|16|16|16|16|16|16|16|16|24|24|24|24|28|28|30|31|16|16|16|16|16|16|16|16|24|24|24|24|28|28|30|31|16|16|16|16|16|16|16|16|24|24|24|24|28|28|30|31|
|17|17|17|17|17|17|17|17|25|25|25|25|29|29|31|31|17|17|17|17|17|17|17|17|25|25|25|25|29|29|31|31|49|49|49|49|49|49|49|49|57|57|57|57|61|61|63|63|49|49|49|49|49|49|49|49|57|57|57|57|61|61|63|63|
|18|19|18|19|18|19|18|19|26|27|26|27|30|31|30|31|18|19|18|19|18|19|18|19|26|27|26|27|30|31|30|31|18|19|18|19|18|19|18|19|26|27|26|27|30|31|30|31|18|19|18|19|18|19|18|19|26|27|26|27|30|31|30|31|
|19|19|19|19|19|19|19|19|27|27|27|27|31|31|31|31|19|19|19|19|19|19|19|19|27|27|27|27|31|31|31|31|51|51|51|51|51|51|51|51|59|59|59|59|63|63|63|63|51|51|51|51|51|51|51|51|59|59|59|59|63|63|63|63|
|20|20|22|23|20|20|22|23|28|28|30|31|28|28|30|31|20|20|22|23|20|20|22|23|28|28|30|31|28|28|30|31|20|20|22|23|20|20|22|23|28|28|30|31|28|28|30|31|20|20|22|23|20|20|22|23|28|28|30|31|28|28|30|31|
|21|21|23|23|21|21|23|23|29|29|31|31|29|29|31|31|21|21|23|23|21|21|23|23|29|29|31|31|29|29|31|31|53|53|55|55|53|53|55|55|61|61|63|63|61|61|63|63|53|53|55|55|53|53|55|55|61|61|63|63|61|61|63|63|
|22|23|22|23|22|23|22|23|30|31|30|31|30|31|30|31|22|23|22|23|22|23|22|23|30|31|30|31|30|31|30|31|22|23|22|23|22|23|22|23|30|31|30|31|30|31|30|31|22|23|22|23|22|23|22|23|30|31|30|31|30|31|30|31|
|23|23|23|23|23|23|23|23|31|31|31|31|31|31|31|31|23|23|23|23|23|23|23|23|31|31|31|31|31|31|31|31|55|55|55|55|55|55|55|55|63|63|63|63|63|63|63|63|55|55|55|55|55|55|55|55|63|63|63|63|63|63|63|63|
|24|24|24|24|28|28|30|31|24|24|24|24|28|28|30|31|24|24|24|24|28|28|30|31|24|24|24|24|28|28|30|31|24|24|24|24|28|28|30|31|24|24|24|24|28|28|30|31|24|24|24|24|28|28|30|31|24|24|24|24|28|28|30|31|
|25|25|25|25|29|29|31|31|25|25|25|25|29|29|31|31|25|25|25|25|29|29|31|31|25|25|25|25|29|29|31|31|57|57|57|57|61|61|63|63|57|57|57|57|61|61|63|63|57|57|57|57|61|61|63|63|57|57|57|57|61|61|63|63|
|26|27|26|27|30|31|30|31|26|27|26|27|30|31|30|31|26|27|26|27|30|31|30|31|26|27|26|27|30|31|30|31|26|27|26|27|30|31|30|31|26|27|26|27|30|31|30|31|26|27|26|27|30|31|30|31|26|27|26|27|30|31|30|31|
|27|27|27|27|31|31|31|31|27|27|27|27|31|31|31|31|27|27|27|27|31|31|31|31|27|27|27|27|31|31|31|31|59|59|59|59|63|63|63|63|59|59|59|59|63|63|63|63|59|59|59|59|63|63|63|63|59|59|59|59|63|63|63|63|
|28|28|30|31|28|28|30|31|28|28|30|31|28|28|30|31|28|28|30|31|28|28|30|31|28|28|30|31|28|28|30|31|28|28|30|31|28|28|30|31|28|28|30|31|28|28|30|31|28|28|30|31|28|28|30|31|28|28|30|31|28|28|30|31|
|29|29|31|31|29|29|31|31|29|29|31|31|29|29|31|31|29|29|31|31|29|29|31|31|29|29|31|31|29|29|31|31|61|61|63|63|61|61|63|63|61|61|63|63|61|61|63|63|61|61|63|63|61|61|63|63|61|61|63|63|61|61|63|63|
|30|31|30|31|30|31|30|31|30|31|30|31|30|31|30|31|30|31|30|31|30|31|30|31|30|31|30|31|30|31|30|31|30|31|30|31|30|31|30|31|30|31|30|31|30|31|30|31|30|31|30|31|30|31|30|31|30|31|30|31|30|31|30|31|
|31|31|31|31|31|31|31|31|31|31|31|31|31|31|31|31|31|31|31|31|31|31|31|31|31|31|31|31|31|31|31|31|63|63|63|63|63|63|63|63|63|63|63|63|63|63|63|63|63|63|63|63|63|63|63|63|63|63|63|63|63|63|63|63|
|32|32|32|32|32|32|32|32|32|32|32|32|32|32|32|32|48|48|48|48|48|48|48|48|56|56|56|56|60|60|62|63|32|32|32|32|32|32|32|32|32|32|32|32|32|32|32|32|48|48|48|48|48|48|48|48|56|56|56|56|60|60|62|63|
|33|33|33|33|33|33|33|33|33|33|33|33|33|33|33|33|49|49|49|49|49|49|49|49|57|57|57|57|61|61|63|63|33|33|33|33|33|33|33|33|33|33|33|33|33|33|33|33|49|49|49|49|49|49|49|49|57|57|57|57|61|61|63|63|
|34|35|34|35|34|35|34|35|34|35|34|35|34|35|34|35|50|51|50|51|50|51|50|51|58|59|58|59|62|63|62|63|34|35|34|35|34|35|34|35|34|35|34|35|34|35|34|35|50|51|50|51|50|51|50|51|58|59|58|59|62|63|62|63|
|35|35|35|35|35|35|35|35|35|35|35|35|35|35|35|35|51|51|51|51|51|51|51|51|59|59|59|59|63|63|63|63|35|35|35|35|35|35|35|35|35|35|35|35|35|35|35|35|51|51|51|51|51|51|51|51|59|59|59|59|63|63|63|63|
|36|36|38|39|36|36|38|39|36|36|38|39|36|36|38|39|52|52|54|55|52|52|54|55|60|60|62|63|60|60|62|63|36|36|38|39|36|36|38|39|36|36|38|39|36|36|38|39|52|52|54|55|52|52|54|55|60|60|62|63|60|60|62|63|
|37|37|39|39|37|37|39|39|37|37|39|39|37|37|39|39|53|53|55|55|53|53|55|55|61|61|63|63|61|61|63|63|37|37|39|39|37|37|39|39|37|37|39|39|37|37|39|39|53|53|55|55|53|53|55|55|61|61|63|63|61|61|63|63|
|38|39|38|39|38|39|38|39|38|39|38|39|38|39|38|39|54|55|54|55|54|55|54|55|62|63|62|63|62|63|62|63|38|39|38|39|38|39|38|39|38|39|38|39|38|39|38|39|54|55|54|55|54|55|54|55|62|63|62|63|62|63|62|63|
|39|39|39|39|39|39|39|39|39|39|39|39|39|39|39|39|55|55|55|55|55|55|55|55|63|63|63|63|63|63|63|63|39|39|39|39|39|39|39|39|39|39|39|39|39|39|39|39|55|55|55|55|55|55|55|55|63|63|63|63|63|63|63|63|
|40|40|40|40|44|44|46|47|40|40|40|40|44|44|46|47|56|56|56|56|60|60|62|63|56|56|56|56|60|60|62|63|40|40|40|40|44|44|46|47|40|40|40|40|44|44|46|47|56|56|56|56|60|60|62|63|56|56|56|56|60|60|62|63|
|41|41|41|41|45|45|47|47|41|41|41|41|45|45|47|47|57|57|57|57|61|61|63|63|57|57|57|57|61|61|63|63|41|41|41|41|45|45|47|47|41|41|41|41|45|45|47|47|57|57|57|57|61|61|63|63|57|57|57|57|61|61|63|63|
|42|43|42|43|46|47|46|47|42|43|42|43|46|47|46|47|58|59|58|59|62|63|62|63|58|59|58|59|62|63|62|63|42|43|42|43|46|47|46|47|42|43|42|43|46|47|46|47|58|59|58|59|62|63|62|63|58|59|58|59|62|63|62|63|
|43|43|43|43|47|47|47|47|43|43|43|43|47|47|47|47|59|59|59|59|63|63|63|63|59|59|59|59|63|63|63|63|43|43|43|43|47|47|47|47|43|43|43|43|47|47|47|47|59|59|59|59|63|63|63|63|59|59|59|59|63|63|63|63|
|44|44|46|47|44|44|46|47|44|44|46|47|44|44|46|47|60|60|62|63|60|60|62|63|60|60|62|63|60|60|62|63|44|44|46|47|44|44|46|47|44|44|46|47|44|44|46|47|60|60|62|63|60|60|62|63|60|60|62|63|60|60|62|63|
|45|45|47|47|45|45|47|47|45|45|47|47|45|45|47|47|61|61|63|63|61|61|63|63|61|61|63|63|61|61|63|63|45|45|47|47|45|45|47|47|45|45|47|47|45|45|47|47|61|61|63|63|61|61|63|63|61|61|63|63|61|61|63|63|
|46|47|46|47|46|47|46|47|46|47|46|47|46|47|46|47|62|63|62|63|62|63|62|63|62|63|62|63|62|63|62|63|46|47|46|47|46|47|46|47|46|47|46|47|46|47|46|47|62|63|62|63|62|63|62|63|62|63|62|63|62|63|62|63|
|47|47|47|47|47|47|47|47|47|47|47|47|47|47|47|47|63|63|63|63|63|63|63|63|63|63|63|63|63|63|63|63|47|47|47|47|47|47|47|47|47|47|47|47|47|47|47|47|63|63|63|63|63|63|63|63|63|63|63|63|63|63|63|63|
|48|48|48|48|48|48|48|48|56|56|56|56|60|60|62|63|48|48|48|48|48|48|48|48|56|56|56|56|60|60|62|63|48|48|48|48|48|48|48|48|56|56|56|56|60|60|62|63|48|48|48|48|48|48|48|48|56|56|56|56|60|60|62|63|
|49|49|49|49|49|49|49|49|57|57|57|57|61|61|63|63|49|49|49|49|49|49|49|49|57|57|57|57|61|61|63|63|49|49|49|49|49|49|49|49|57|57|57|57|61|61|63|63|49|49|49|49|49|49|49|49|57|57|57|57|61|61|63|63|
|50|51|50|51|50|51|50|51|58|59|58|59|62|63|62|63|50|51|50|51|50|51|50|51|58|59|58|59|62|63|62|63|50|51|50|51|50|51|50|51|58|59|58|59|62|63|62|63|50|51|50|51|50|51|50|51|58|59|58|59|62|63|62|63|
|51|51|51|51|51|51|51|51|59|59|59|59|63|63|63|63|51|51|51|51|51|51|51|51|59|59|59|59|63|63|63|63|51|51|51|51|51|51|51|51|59|59|59|59|63|63|63|63|51|51|51|51|51|51|51|51|59|59|59|59|63|63|63|63|
|52|52|54|55|52|52|54|55|60|60|62|63|60|60|62|63|52|52|54|55|52|52|54|55|60|60|62|63|60|60|62|63|52|52|54|55|52|52|54|55|60|60|62|63|60|60|62|63|52|52|54|55|52|52|54|55|60|60|62|63|60|60|62|63|
|53|53|55|55|53|53|55|55|61|61|63|63|61|61|63|63|53|53|55|55|53|53|55|55|61|61|63|63|61|61|63|63|53|53|55|55|53|53|55|55|61|61|63|63|61|61|63|63|53|53|55|55|53|53|55|55|61|61|63|63|61|61|63|63|
|54|55|54|55|54|55|54|55|62|63|62|63|62|63|62|63|54|55|54|55|54|55|54|55|62|63|62|63|62|63|62|63|54|55|54|55|54|55|54|55|62|63|62|63|62|63|62|63|54|55|54|55|54|55|54|55|62|63|62|63|62|63|62|63|
|55|55|55|55|55|55|55|55|63|63|63|63|63|63|63|63|55|55|55|55|55|55|55|55|63|63|63|63|63|63|63|63|55|55|55|55|55|55|55|55|63|63|63|63|63|63|63|63|55|55|55|55|55|55|55|55|63|63|63|63|63|63|63|63|
|56|56|56|56|60|60|62|63|56|56|56|56|60|60|62|63|56|56|56|56|60|60|62|63|56|56|56|56|60|60|62|63|56|56|56|56|60|60|62|63|56|56|56|56|60|60|62|63|56|56|56|56|60|60|62|63|56|56|56|56|60|60|62|63|
|57|57|57|57|61|61|63|63|57|57|57|57|61|61|63|63|57|57|57|57|61|61|63|63|57|57|57|57|61|61|63|63|57|57|57|57|61|61|63|63|57|57|57|57|61|61|63|63|57|57|57|57|61|61|63|63|57|57|57|57|61|61|63|63|
|58|59|58|59|62|63|62|63|58|59|58|59|62|63|62|63|58|59|58|59|62|63|62|63|58|59|58|59|62|63|62|63|58|59|58|59|62|63|62|63|58|59|58|59|62|63|62|63|58|59|58|59|62|63|62|63|58|59|58|59|62|63|62|63|
|59|59|59|59|63|63|63|63|59|59|59|59|63|63|63|63|59|59|59|59|63|63|63|63|59|59|59|59|63|63|63|63|59|59|59|59|63|63|63|63|59|59|59|59|63|63|63|63|59|59|59|59|63|63|63|63|59|59|59|59|63|63|63|63|
|60|60|62|63|60|60|62|63|60|60|62|63|60|60|62|63|60|60|62|63|60|60|62|63|60|60|62|63|60|60|62|63|60|60|62|63|60|60|62|63|60|60|62|63|60|60|62|63|60|60|62|63|60|60|62|63|60|60|62|63|60|60|62|63|
|61|61|63|63|61|61|63|63|61|61|63|63|61|61|63|63|61|61|63|63|61|61|63|63|61|61|63|63|61|61|63|63|61|61|63|63|61|61|63|63|61|61|63|63|61|61|63|63|61|61|63|63|61|61|63|63|61|61|63|63|61|61|63|63|
|62|63|62|63|62|63|62|63|62|63|62|63|62|63|62|63|62|63|62|63|62|63|62|63|62|63|62|63|62|63|62|63|62|63|62|63|62|63|62|63|62|63|62|63|62|63|62|63|62|63|62|63|62|63|62|63|62|63|62|63|62|63|62|63|
|63|63|63|63|63|63|63|63|63|63|63|63|63|63|63|63|63|63|63|63|63|63|63|63|63|63|63|63|63|63|63|63|63|63|63|63|63|63|63|63|63|63|63|63|63|63|63|63|63|63|63|63|63|63|63|63|63|63|63|63|63|63|63|63|

Figure 8: Matrix $M_{L_1}$ of composition $\mathbf{x} \circ \mathbf{y}$ implemented by $L_1$. Symmetric ($m_{ij} = m_{ji}$) entries are underlined.

| | | | | | | | | | | | | | | | | | | | | | | | | | | | | | | | | | | | | | | | | | | | | | | | | | | | | | | | | | | | | | | | |
|---|---|---|---|---|---|---|---|---|---|---|---|---|---|---|---|---|---|---|---|---|---|---|---|---|---|---|---|---|---|---|---|---|---|---|---|---|---|---|---|---|---|---|---|---|---|---|---|---|---|---|---|---|---|---|---|---|---|---|---|---|---|---|---|
| 0 | 1 | 2 | 2 | 4 | 5 | 4 | 5 | 8 | 9 | 10 | 10 | 8 | 9 | 10 | 10 | 16 | 17 | 18 | 18 | 20 | 21 | 20 | 21 | 16 | 17 | 18 | 18 | 20 | 21 | 20 | 21 | 32 | 33 | 34 | 34 | 36 | 37 | 36 | 37 | 40 | 41 | 42 | 42 | 40 | 41 | 42 | 42 | 32 | 33 | 34 | 34 | 36 | 37 | 36 | 37 | 40 | 41 | 42 | 42 | 40 | 41 | 42 | 42 |
| 1 | 1 | 3 | 3 | 5 | 5 | 5 | 5 | 9 | 9 | 11 | 11 | 9 | 9 | 11 | 11 | 17 | 17 | 19 | 19 | 21 | 21 | 21 | 21 | 17 | 17 | 19 | 19 | 21 | 21 | 21 | 21 | 1 | 1 | 3 | 3 | 5 | 5 | 5 | 5 | 9 | 9 | 11 | 11 | 9 | 9 | 11 | 11 | 17 | 17 | 19 | 19 | 21 | 21 | 21 | 21 | 17 | 17 | 19 | 19 | 21 | 21 | 21 | 21 |
| 2 | 2 | 2 | 2 | 6 | 6 | 6 | 6 | 10 | 10 | 10 | 10 | 10 | 10 | 10 | 10 | 18 | 18 | 18 | 18 | 22 | 22 | 22 | 22 | 18 | 18 | 18 | 18 | 22 | 22 | 22 | 22 | 34 | 34 | 34 | 34 | 38 | 38 | 38 | 38 | 42 | 42 | 42 | 42 | 42 | 42 | 42 | 42 | 34 | 34 | 34 | 34 | 38 | 38 | 38 | 38 | 42 | 42 | 42 | 42 | 42 | 42 | 42 | 42 |
| 3 | 3 | 3 | 3 | 7 | 7 | 7 | 7 | 11 | 11 | 11 | 11 | 11 | 11 | 11 | 11 | 19 | 19 | 19 | 19 | 23 | 23 | 23 | 23 | 19 | 19 | 19 | 19 | 23 | 23 | 23 | 23 | 3 | 3 | 3 | 3 | 7 | 7 | 7 | 7 | 11 | 11 | 11 | 11 | 11 | 11 | 11 | 11 | 19 | 19 | 19 | 19 | 23 | 23 | 23 | 23 | 19 | 19 | 19 | 19 | 23 | 23 | 23 | 23 |
| 4 | 5 | 4 | 5 | 4 | 5 | 4 | 5 | 12 | 13 | 12 | 13 | 12 | 13 | 12 | 13 | 20 | 21 | 20 | 21 | 20 | 21 | 20 | 21 | 20 | 21 | 20 | 21 | 20 | 21 | 20 | 21 | 36 | 37 | 36 | 37 | 36 | 37 | 36 | 37 | 44 | 45 | 44 | 45 | 44 | 45 | 44 | 45 | 36 | 37 | 36 | 37 | 36 | 37 | 36 | 37 | 44 | 45 | 44 | 45 | 44 | 45 | 44 | 45 |
| 5 | 5 | 5 | 5 | 5 | 5 | 5 | 5 | 13 | 13 | 13 | 13 | 13 | 13 | 13 | 13 | 21 | 21 | 21 | 21 | 21 | 21 | 21 | 21 | 21 | 21 | 21 | 21 | 21 | 21 | 21 | 21 | 5 | 5 | 5 | 5 | 5 | 5 | 5 | 5 | 13 | 13 | 13 | 13 | 13 | 13 | 13 | 13 | 21 | 21 | 21 | 21 | 21 | 21 | 21 | 21 | 21 | 21 | 21 | 21 | 21 | 21 | 21 | 21 |
| 6 | 6 | 6 | 6 | 6 | 6 | 6 | 6 | 14 | 14 | 14 | 14 | 14 | 14 | 14 | 14 | 22 | 22 | 22 | 22 | 22 | 22 | 22 | 22 | 22 | 22 | 22 | 22 | 22 | 22 | 22 | 22 | 38 | 38 | 38 | 38 | 38 | 38 | 38 | 38 | 46 | 46 | 46 | 46 | 46 | 46 | 46 | 46 | 38 | 38 | 38 | 38 | 38 | 38 | 38 | 38 | 46 | 46 | 46 | 46 | 46 | 46 | 46 | 46 |
| 7 | 7 | 7 | 7 | 7 | 7 | 7 | 7 | 15 | 15 | 15 | 15 | 15 | 15 | 15 | 15 | 23 | 23 | 23 | 23 | 23 | 23 | 23 | 23 | 23 | 23 | 23 | 23 | 23 | 23 | 23 | 23 | 7 | 7 | 7 | 7 | 7 | 7 | 7 | 7 | 15 | 15 | 15 | 15 | 15 | 15 | 15 | 15 | 23 | 23 | 23 | 23 | 23 | 23 | 23 | 23 | 23 | 23 | 23 | 23 | 23 | 23 | 23 | 23 |
| 8 | 9 | 10 | 10 | 8 | 9 | 10 | 10 | 8 | 9 | 10 | 10 | 8 | 9 | 10 | 10 | 24 | 25 | 26 | 26 | 24 | 25 | 26 | 26 | 24 | 25 | 26 | 26 | 24 | 25 | 26 | 26 | 40 | 41 | 42 | 42 | 40 | 41 | 42 | 42 | 40 | 41 | 42 | 42 | 40 | 41 | 42 | 42 | 40 | 41 | 42 | 42 | 40 | 41 | 42 | 42 | 40 | 41 | 42 | 42 | 40 | 41 | 42 | 42 |
| 9 | 9 | 11 | 11 | 9 | 9 | 11 | 11 | 9 | 9 | 11 | 11 | 9 | 9 | 11 | 11 | 25 | 25 | 27 | 27 | 25 | 25 | 27 | 27 | 25 | 25 | 27 | 27 | 25 | 25 | 27 | 27 | 9 | 9 | 11 | 11 | 9 | 9 | 11 | 11 | 9 | 9 | 11 | 11 | 9 | 9 | 11 | 11 | 25 | 25 | 27 | 27 | 25 | 25 | 27 | 27 | 25 | 25 | 27 | 27 | 25 | 25 | 27 | 27 |
| 10 | 10 | 10 | 10 | 10 | 10 | 10 | 10 | 10 | 10 | 10 | 10 | 10 | 10 | 10 | 10 | 26 | 26 | 26 | 26 | 26 | 26 | 26 | 26 | 26 | 26 | 26 | 26 | 26 | 26 | 26 | 26 | 42 | 42 | 42 | 42 | 42 | 42 | 42 | 42 | 42 | 42 | 42 | 42 | 42 | 42 | 42 | 42 | 42 | 42 | 42 | 42 | 42 | 42 | 42 | 42 | 42 | 42 | 42 | 42 | 42 | 42 | 42 | 42 |
| 11 | 11 | 11 | 11 | 11 | 11 | 11 | 11 | 11 | 11 | 11 | 11 | 11 | 11 | 11 | 11 | 27 | 27 | 27 | 27 | 27 | 27 | 27 | 27 | 27 | 27 | 27 | 27 | 27 | 27 | 27 | 27 | 11 | 11 | 11 | 11 | 11 | 11 | 11 | 11 | 11 | 11 | 11 | 11 | 11 | 11 | 11 | 11 | 27 | 27 | 27 | 27 | 27 | 27 | 27 | 27 | 27 | 27 | 27 | 27 | 27 | 27 | 27 | 27 |
| 12 | 13 | 12 | 13 | 12 | 13 | 12 | 13 | 12 | 13 | 12 | 13 | 12 | 13 | 12 | 13 | 28 | 29 | 28 | 29 | 28 | 29 | 28 | 29 | 28 | 29 | 28 | 29 | 28 | 29 | 28 | 29 | 44 | 45 | 44 | 45 | 44 | 45 | 44 | 45 | 44 | 45 | 44 | 45 | 44 | 45 | 44 | 45 | 44 | 45 | 44 | 45 | 44 | 45 | 44 | 45 | 44 | 45 | 44 | 45 | 44 | 45 | 44 | 45 |
| 13 | 13 | 13 | 13 | 13 | 13 | 13 | 13 | 13 | 13 | 13 | 13 | 13 | 13 | 13 | 13 | 29 | 29 | 29 | 29 | 29 | 29 | 29 | 29 | 29 | 29 | 29 | 29 | 29 | 29 | 29 | 29 | 13 | 13 | 13 | 13 | 13 | 13 | 13 | 13 | 13 | 13 | 13 | 13 | 13 | 13 | 13 | 13 | 29 | 29 | 29 | 29 | 29 | 29 | 29 | 29 | 29 | 29 | 29 | 29 | 29 | 29 | 29 | 29 |
| 14 | 14 | 14 | 14 | 14 | 14 | 14 | 14 | 14 | 14 | 14 | 14 | 14 | 14 | 14 | 14 | 30 | 30 | 30 | 30 | 30 | 30 | 30 | 30 | 30 | 30 | 30 | 30 | 30 | 30 | 30 | 30 | 46 | 46 | 46 | 46 | 46 | 46 | 46 | 46 | 46 | 46 | 46 | 46 | 46 | 46 | 46 | 46 | 46 | 46 | 46 | 46 | 46 | 46 | 46 | 46 | 46 | 46 | 46 | 46 | 46 | 46 | 46 | 46 |
| 15 | 15 | 15 | 15 | 15 | 15 | 15 | 15 | 15 | 15 | 15 | 15 | 15 | 15 | 15 | 15 | 31 | 31 | 31 | 31 | 31 | 31 | 31 | 31 | 31 | 31 | 31 | 31 | 31 | 31 | 31 | 31 | 15 | 15 | 15 | 15 | 15 | 15 | 15 | 15 | 15 | 15 | 15 | 15 | 15 | 15 | 15 | 15 | 31 | 31 | 31 | 31 | 31 | 31 | 31 | 31 | 31 | 31 | 31 | 31 | 31 | 31 | 31 | 31 |
| 16 | 17 | 18 | 18 | 20 | 21 | 20 | 21 | 16 | 17 | 18 | 18 | 20 | 21 | 20 | 21 | 16 | 17 | 18 | 18 | 20 | 21 | 20 | 21 | 16 | 17 | 18 | 18 | 20 | 21 | 20 | 21 | 48 | 49 | 50 | 50 | 52 | 53 | 52 | 53 | 48 | 49 | 50 | 50 | 52 | 53 | 52 | 53 | 48 | 49 | 50 | 50 | 52 | 53 | 52 | 53 | 48 | 49 | 50 | 50 | 52 | 53 | 52 | 53 |
| 17 | 17 | 19 | 19 | 21 | 21 | 21 | 21 | 17 | 17 | 19 | 19 | 21 | 21 | 21 | 21 | 17 | 17 | 19 | 19 | 21 | 21 | 21 | 21 | 17 | 17 | 19 | 19 | 21 | 21 | 21 | 21 | 17 | 17 | 19 | 19 | 21 | 21 | 21 | 21 | 17 | 17 | 19 | 19 | 21 | 21 | 21 | 21 | 17 | 17 | 19 | 19 | 21 | 21 | 21 | 21 | 17 | 17 | 19 | 19 | 21 | 21 | 21 | 21 |
| 18 | 18 | 18 | 18 | 22 | 22 | 22 | 22 | 18 | 18 | 18 | 18 | 22 | 22 | 22 | 22 | 18 | 18 | 18 | 18 | 22 | 22 | 22 | 22 | 18 | 18 | 18 | 18 | 22 | 22 | 22 | 22 | 50 | 50 | 50 | 50 | 54 | 54 | 54 | 54 | 50 | 50 | 50 | 50 | 54 | 54 | 54 | 54 | 50 | 50 | 50 | 50 | 54 | 54 | 54 | 54 | 50 | 50 | 50 | 50 | 54 | 54 | 54 | 54 |
| 19 | 19 | 19 | 19 | 23 | 23 | 23 | 23 | 19 | 19 | 19 | 19 | 23 | 23 | 23 | 23 | 19 | 19 | 19 | 19 | 23 | 23 | 23 | 23 | 19 | 19 | 19 | 19 | 23 | 23 | 23 | 23 | 19 | 19 | 19 | 19 | 23 | 23 | 23 | 23 | 19 | 19 | 19 | 19 | 23 | 23 | 23 | 23 | 19 | 19 | 19 | 19 | 23 | 23 | 23 | 23 | 19 | 19 | 19 | 19 | 23 | 23 | 23 | 23 |
| 20 | 21 | 20 | 21 | 20 | 21 | 20 | 21 | 20 | 21 | 20 | 21 | 20 | 21 | 20 | 21 | 20 | 21 | 20 | 21 | 20 | 21 | 20 | 21 | 20 | 21 | 20 | 21 | 20 | 21 | 20 | 21 | 52 | 53 | 52 | 53 | 52 | 53 | 52 | 53 | 52 | 53 | 52 | 53 | 52 | 53 | 52 | 53 | 52 | 53 | 52 | 53 | 52 | 53 | 52 | 53 | 52 | 53 | 52 | 53 | 52 | 53 | 52 | 53 |
| 21 | 21 | 21 | 21 | 21 | 21 | 21 | 21 | 21 | 21 | 21 | 21 | 21 | 21 | 21 | 21 | 21 | 21 | 21 | 21 | 21 | 21 | 21 | 21 | 21 | 21 | 21 | 21 | 21 | 21 | 21 | 21 | 21 | 21 | 21 | 21 | 21 | 21 | 21 | 21 | 21 | 21 | 21 | 21 | 21 | 21 | 21 | 21 | 21 | 21 | 21 | 21 | 21 | 21 | 21 | 21 | 21 | 21 | 21 | 21 | 21 | 21 | 21 | 21 |
| 22 | 22 | 22 | 22 | 22 | 22 | 22 | 22 | 22 | 22 | 22 | 22 | 22 | 22 | 22 | 22 | 22 | 22 | 22 | 22 | 22 | 22 | 22 | 22 | 22 | 22 | 22 | 22 | 22 | 22 | 22 | 22 | 54 | 54 | 54 | 54 | 54 | 54 | 54 | 54 | 54 | 54 | 54 | 54 | 54 | 54 | 54 | 54 | 54 | 54 | 54 | 54 | 54 | 54 | 54 | 54 | 54 | 54 | 54 | 54 | 54 | 54 | 54 | 54 |
| 23 | 23 | 23 | 23 | 23 | 23 | 23 | 23 | 23 | 23 | 23 | 23 | 23 | 23 | 23 | 23 | 23 | 23 | 23 | 23 | 23 | 23 | 23 | 23 | 23 | 23 | 23 | 23 | 23 | 23 | 23 | 23 | 23 | 23 | 23 | 23 | 23 | 23 | 23 | 23 | 23 | 23 | 23 | 23 | 23 | 23 | 23 | 23 | 23 | 23 | 23 | 23 | 23 | 23 | 23 | 23 | 23 | 23 | 23 | 23 | 23 | 23 | 23 | 23 |
| 24 | 25 | 26 | 26 | 24 | 25 | 26 | 26 | 24 | 25 | 26 | 26 | 24 | 25 | 26 | 26 | 24 | 25 | 26 | 26 | 24 | 25 | 26 | 26 | 24 | 25 | 26 | 26 | 24 | 25 | 26 | 26 | 56 | 57 | 58 | 58 | 56 | 57 | 58 | 58 | 56 | 57 | 58 | 58 | 56 | 57 | 58 | 58 | 56 | 57 | 58 | 58 | 56 | 57 | 58 | 58 | 56 | 57 | 58 | 58 | 56 | 57 | 58 | 58 |
| 25 | 25 | 27 | 27 | 25 | 25 | 27 | 27 | 25 | 25 | 27 | 27 | 25 | 25 | 27 | 27 | 25 | 25 | 27 | 27 | 25 | 25 | 27 | 27 | 25 | 25 | 27 | 27 | 25 | 25 | 27 | 27 | 25 | 25 | 27 | 27 | 25 | 25 | 27 | 27 | 25 | 25 | 27 | 27 | 25 | 25 | 27 | 27 | 25 | 25 | 27 | 27 | 25 | 25 | 27 | 27 | 25 | 25 | 27 | 27 | 25 | 25 | 27 | 27 |
| 26 | 26 | 26 | 26 | 26 | 26 | 26 | 26 | 26 | 26 | 26 | 26 | 26 | 26 | 26 | 26 | 26 | 26 | 26 | 26 | 26 | 26 | 26 | 26 | 26 | 26 | 26 | 26 | 26 | 26 | 26 | 26 | 58 | 58 | 58 | 58 | 58 | 58 | 58 | 58 | 58 | 58 | 58 | 58 | 58 | 58 | 58 | 58 | 58 | 58 | 58 | 58 | 58 | 58 | 58 | 58 | 58 | 58 | 58 | 58 | 58 | 58 | 58 | 58 |
| 27 | 27 | 27 | 27 | 27 | 27 | 27 | 27 | 27 | 27 | 27 | 27 | 27 | 27 | 27 | 27 | 27 | 27 | 27 | 27 | 27 | 27 | 27 | 27 | 27 | 27 | 27 | 27 | 27 | 27 | 27 | 27 | 27 | 27 | 27 | 27 | 27 | 27 | 27 | 27 | 27 | 27 | 27 | 27 | 27 | 27 | 27 | 27 | 27 | 27 | 27 | 27 | 27 | 27 | 27 | 27 | 27 | 27 | 27 | 27 | 27 | 27 | 27 | 27 |
| 28 | 29 | 28 | 29 | 28 | 29 | 28 | 29 | 28 | 29 | 28 | 29 | 28 | 29 | 28 | 29 | 28 | 29 | 28 | 29 | 28 | 29 | 28 | 29 | 28 | 29 | 28 | 29 | 28 | 29 | 28 | 29 | 60 | 61 | 60 | 61 | 60 | 61 | 60 | 61 | 60 | 61 | 60 | 61 | 60 | 61 | 60 | 61 | 60 | 61 | 60 | 61 | 60 | 61 | 60 | 61 | 60 | 61 | 60 | 61 | 60 | 61 | 60 | 61 |
| 29 | 29 | 29 | 29 | 29 | 29 | 29 | 29 | 29 | 29 | 29 | 29 | 29 | 29 | 29 | 29 | 29 | 29 | 29 | 29 | 29 | 29 | 29 | 29 | 29 | 29 | 29 | 29 | 29 | 29 | 29 | 29 | 29 | 29 | 29 | 29 | 29 | 29 | 29 | 29 | 29 | 29 | 29 | 29 | 29 | 29 | 29 | 29 | 29 | 29 | 29 | 29 | 29 | 29 | 29 | 29 | 29 | 29 | 29 | 29 | 29 | 29 | 29 | 29 |
| 30 | 30 | 30 | 30 | 30 | 30 | 30 | 30 | 30 | 30 | 30 | 30 | 30 | 30 | 30 | 30 | 30 | 30 | 30 | 30 | 30 | 30 | 30 | 30 | 30 | 30 | 30 | 30 | 30 | 30 | 30 | 30 | 62 | 62 | 62 | 62 | 62 | 62 | 62 | 62 | 62 | 62 | 62 | 62 | 62 | 62 | 62 | 62 | 62 | 62 | 62 | 62 | 62 | 62 | 62 | 62 | 62 | 62 | 62 | 62 | 62 | 62 | 62 | 62 |
| 31 | 31 | 31 | 31 | 31 | 31 | 31 | 31 | 31 | 31 | 31 | 31 | 31 | 31 | 31 | 31 | 31 | 31 | 31 | 31 | 31 | 31 | 31 | 31 | 31 | 31 | 31 | 31 | 31 | 31 | 31 | 31 | 31 | 31 | 31 | 31 | 31 | 31 | 31 | 31 | 31 | 31 | 31 | 31 | 31 | 31 | 31 | 31 | 31 | 31 | 31 | 31 | 31 | 31 | 31 | 31 | 31 | 31 | 31 | 31 | 31 | 31 | 31 | 31 |
| 32 | 33 | 34 | 34 | 36 | 37 | 36 | 37 | 40 | 41 | 42 | 42 | 40 | 41 | 42 | 42 | 32 | 33 | 34 | 34 | 36 | 37 | 36 | 37 | 40 | 41 | 42 | 42 | 40 | 41 | 42 | 42 | 32 | 33 | 34 | 34 | 36 | 37 | 36 | 37 | 40 | 41 | 42 | 42 | 40 | 41 | 42 | 42 | 32 | 33 | 34 | 34 | 36 | 37 | 36 | 37 | 40 | 41 | 42 | 42 | 40 | 41 | 42 | 42 |
| 33 | 33 | 35 | 35 | 37 | 37 | 37 | 37 | 41 | 41 | 43 | 43 | 41 | 41 | 43 | 43 | 33 | 33 | 35 | 35 | 37 | 37 | 37 | 37 | 41 | 41 | 43 | 43 | 41 | 41 | 43 | 43 | 33 | 33 | 35 | 35 | 37 | 37 | 37 | 37 | 41 | 41 | 43 | 43 | 41 | 41 | 43 | 43 | 33 | 33 | 35 | 35 | 37 | 37 | 37 | 37 | 41 | 41 | 43 | 43 | 41 | 41 | 43 | 43 |
| 34 | 34 | 34 | 34 | 38 | 38 | 38 | 38 | 42 | 42 | 42 | 42 | 42 | 42 | 42 | 42 | 34 | 34 | 34 | 34 | 38 | 38 | 38 | 38 | 42 | 42 | 42 | 42 | 42 | 42 | 42 | 42 | 34 | 34 | 34 | 34 | 38 | 38 | 38 | 38 | 42 | 42 | 42 | 42 | 42 | 42 | 42 | 42 | 34 | 34 | 34 | 34 | 38 | 38 | 38 | 38 | 42 | 42 | 42 | 42 | 42 | 42 | 42 | 42 |
| 35 | 35 | 35 | 35 | 39 | 39 | 39 | 39 | 43 | 43 | 43 | 43 | 43 | 43 | 43 | 43 | 35 | 35 | 35 | 35 | 39 | 39 | 39 | 39 | 43 | 43 | 43 | 43 | 43 | 43 | 43 | 43 | 35 | 35 | 35 | 35 | 39 | 39 | 39 | 39 | 43 | 43 | 43 | 43 | 43 | 43 | 43 | 43 | 35 | 35 | 35 | 35 | 39 | 39 | 39 | 39 | 43 | 43 | 43 | 43 | 43 | 43 | 43 | 43 |
| 36 | 37 | 36 | 37 | 36 | 37 | 36 | 37 | 44 | 45 | 44 | 45 | 44 | 45 | 44 | 45 | 36 | 37 | 36 | 37 | 36 | 37 | 36 | 37 | 44 | 45 | 44 | 45 | 44 | 45 | 44 | 45 | 36 | 37 | 36 | 37 | 36 | 37 | 36 | 37 | 44 | 45 | 44 | 45 | 44 | 45 | 44 | 45 | 36 | 37 | 36 | 37 | 36 | 37 | 36 | 37 | 44 | 45 | 44 | 45 | 44 | 45 | 44 | 45 |
| 37 | 37 | 37 | 37 | 37 | 37 | 37 | 37 | 45 | 45 | 45 | 45 | 45 | 45 | 45 | 45 | 37 | 37 | 37 | 37 | 37 | 37 | 37 | 37 | 45 | 45 | 45 | 45 | 45 | 45 | 45 | 45 | 37 | 37 | 37 | 37 | 37 | 37 | 37 | 37 | 45 | 45 | 45 | 45 | 45 | 45 | 45 | 45 | 37 | 37 | 37 | 37 | 37 | 37 | 37 | 37 | 45 | 45 | 45 | 45 | 45 | 45 | 45 | 45 |
| 38 | 38 | 38 | 38 | 38 | 38 | 38 | 38 | 46 | 46 | 46 | 46 | 46 | 46 | 46 | 46 | 38 | 38 | 38 | 38 | 38 | 38 | 38 | 38 | 46 | 46 | 46 | 46 | 46 | 46 | 46 | 46 | 38 | 38 | 38 | 38 | 38 | 38 | 38 | 38 | 46 | 46 | 46 | 46 | 46 | 46 | 46 | 46 | 38 | 38 | 38 | 38 | 38 | 38 | 38 | 38 | 46 | 46 | 46 | 46 | 46 | 46 | 46 | 46 |
| 39 | 39 | 39 | 39 | 39 | 39 | 39 | 39 | 47 | 47 | 47 | 47 | 47 | 47 | 47 | 47 | 39 | 39 | 39 | 39 | 39 | 39 | 39 | 39 | 47 | 47 | 47 | 47 | 47 | 47 | 47 | 47 | 39 | 39 | 39 | 39 | 39 | 39 | 39 | 39 | 47 | 47 | 47 | 47 | 47 | 47 | 47 | 47 | 39 | 39 | 39 | 39 | 39 | 39 | 39 | 39 | 47 | 47 | 47 | 47 | 47 | 47 | 47 | 47 |
| 40 | 41 | 42 | 42 | 40 | 41 | 42 | 42 | 40 | 41 | 42 | 42 | 40 | 41 | 42 | 42 | 40 | 41 | 42 | 42 | 40 | 41 | 42 | 42 | 40 | 41 | 42 | 42 | 40 | 41 | 42 | 42 | 40 | 41 | 42 | 42 | 40 | 41 | 42 | 42 | 40 | 41 | 42 | 42 | 40 | 41 | 42 | 42 | 40 | 41 | 42 | 42 | 40 | 41 | 42 | 42 | 40 | 41 | 42 | 42 | 40 | 41 | 42 | 42 |
| 41 | 41 | 43 | 43 | 41 | 41 | 43 | 43 | 41 | 41 | 43 | 43 | 41 | 41 | 43 | 43 | 41 | 41 | 43 | 43 | 41 | 41 | 43 | 43 | 41 | 41 | 43 | 43 | 41 | 41 | 43 | 43 | 41 | 41 | 43 | 43 | 41 | 41 | 43 | 43 | 41 | 41 | 43 | 43 | 41 | 41 | 43 | 43 | 41 | 41 | 43 | 43 | 41 | 41 | 43 | 43 | 41 | 41 | 43 | 43 | 41 | 41 | 43 | 43 |
| 42 | 42 | 42 | 42 | 42 | 42 | 42 | 42 | 42 | 42 | 42 | 42 | 42 | 42 | 42 | 42 | 42 | 42 | 42 | 42 | 42 | 42 | 42 | 42 | 42 | 42 | 42 | 42 | 42 | 42 | 42 | 42 | 42 | 42 | 42 | 42 | 42 | 42 | 42 | 42 | 42 | 42 | 42 | 42 | 42 | 42 | 42 | 42 | 42 | 42 | 42 | 42 | 42 | 42 | 42 | 42 | 42 | 42 | 42 | 42 | 42 | 42 | 42 | 42 |
| 43 | 43 | 43 | 43 | 43 | 43 | 43 | 43 | 43 | 43 | 43 | 43 | 43 | 43 | 43 | 43 | 43 | 43 | 43 | 43 | 43 | 43 | 43 | 43 | 43 | 43 | 43 | 43 | 43 | 43 | 43 | 43 | 43 | 43 | 43 | 43 | 43 | 43 | 43 | 43 | 43 | 43 | 43 | 43 | 43 | 43 | 43 | 43 | 43 | 43 | 43 | 43 | 43 | 43 | 43 | 43 | 43 | 43 | 43 | 43 | 43 | 43 | 43 | 43 |
| 44 | 45 | 44 | 45 | 44 | 45 | 44 | 45 | 44 | 45 | 44 | 45 | 44 | 45 | 44 | 45 | 44 | 45 | 44 | 45 | 44 | 45 | 44 | 45 | 44 | 45 | 44 | 45 | 44 | 45 | 44 | 45 | 44 | 45 | 44 | 45 | 44 | 45 | 44 | 45 | 44 | 45 | 44 | 45 | 44 | 45 | 44 | 45 | 44 | 45 | 44 | 45 | 44 | 45 | 44 | 45 | 44 | 45 | 44 | 45 | 44 | 45 | 44 | 45 |
| 45 | 45 | 45 | 45 | 45 | 45 | 45 | 45 | 45 | 45 | 45 | 45 | 45 | 45 | 45 | 45 | 45 | 45 | 45 | 45 | 45 | 45 | 45 | 45 | 45 | 45 | 45 | 45 | 45 | 45 | 45 | 45 | 45 | 45 | 45 | 45 | 45 | 45 | 45 | 45 | 45 | 45 | 45 | 45 | 45 | 45 | 45 | 45 | 45 | 45 | 45 | 45 | 45 | 45 | 45 | 45 | 45 | 45 | 45 | 45 | 45 | 45 | 45 | 45 |
| 46 | 46 | 46 | 46 | 46 | 46 | 46 | 46 | 46 | 46 | 46 | 46 | 46 | 46 | 46 | 46 | 46 | 46 | 46 | 46 | 46 | 46 | 46 | 46 | 46 | 46 | 46 | 46 | 46 | 46 | 46 | 46 | 46 | 46 | 46 | 46 | 46 | 46 | 46 | 46 | 46 | 46 | 46 | 46 | 46 | 46 | 46 | 46 | 46 | 46 | 46 | 46 | 46 | 46 | 46 | 46 | 46 | 46 | 46 | 46 | 46 | 46 | 46 | 46 |
| 47 | 47 | 47 | 47 | 47 | 47 | 47 | 47 | 47 | 47 | 47 | 47 | 47 | 47 | 47 | 47 | 47 | 47 | 47 | 47 | 47 | 47 | 47 | 47 | 47 | 47 | 47 | 47 | 47 | 47 | 47 | 47 | 47 | 47 | 47 | 47 | 47 | 47 | 47 | 47 | 47 | 47 | 47 | 47 | 47 | 47 | 47 | 47 | 47 | 47 | 47 | 47 | 47 | 47 | 47 | 47 | 47 | 47 | 47 | 47 | 47 | 47 | 47 | 47 |
| 48 | 49 | 50 | 50 | 52 | 53 | 52 | 53 | 48 | 49 | 50 | 50 | 52 | 53 | 52 | 53 | 48 | 49 | 50 | 50 | 52 | 53 | 52 | 53 | 48 | 49 | 50 | 50 | 52 | 53 | 52 | 53 | 48 | 49 | 50 | 50 | 52 | 53 | 52 | 53 | 48 | 49 | 50 | 50 | 52 | 53 | 52 | 53 | 48 | 49 | 50 | 50 | 52 | 53 | 52 | 53 | 48 | 49 | 50 | 50 | 52 | 53 | 52 | 53 |
| 49 | 49 | 51 | 51 | 53 | 53 | 53 | 53 | 49 | 49 | 51 | 51 | 53 | 53 | 53 | 53 | 49 | 49 | 51 | 51 | 53 | 53 | 53 | 53 | 49 | 49 | 51 | 51 | 53 | 53 | 53 | 53 | 49 | 49 | 51 | 51 | 53 | 53 | 53 | 53 | 49 | 49 | 51 | 51 | 53 | 53 | 53 | 53 | 49 | 49 | 51 | 51 | 53 | 53 | 53 | 53 | 49 | 49 | 51 | 51 | 53 | 53 | 53 | 53 |
| 50 | 50 | 50 | 50 | 54 | 54 | 54 | 54 | 50 | 50 | 50 | 50 | 54 | 54 | 54 | 54 | 50 | 50 | 50 | 50 | 54 | 54 | 54 | 54 | 50 | 50 | 50 | 50 | 54 | 54 | 54 | 54 | 50 | 50 | 50 | 50 | 54 | 54 | 54 | 54 | 50 | 50 | 50 | 50 | 54 | 54 | 54 | 54 | 50 | 50 | 50 | 50 | 54 | 54 | 54 | 54 | 50 | 50 | 50 | 50 | 54 | 54 | 54 | 54 |
| 51 | 51 | 51 | 51 | 55 | 55 | 55 | 55 | 51 | 51 | 51 | 51 | 55 | 55 | 55 | 55 | 51 | 51 | 51 | 51 | 55 | 55 | 55 | 55 | 51 | 51 | 51 | 51 | 55 | 55 | 55 | 55 | 51 | 51 | 51 | 51 | 55 | 55 | 55 | 55 | 51 | 51 | 51 | 51 | 55 | 55 | 55 | 55 | 51 | 51 | 51 | 51 | 55 | 55 | 55 | 55 | 51 | 51 | 51 | 51 | 55 | 55 | 55 | 55 |
| 52 | 53 | 52 | 53 | 52 | 53 | 52 | 53 | 52 | 53 | 52 | 53 | 52 | 53 | 52 | 53 | 52 | 53 | 52 | 53 | 52 | 53 | 52 | 53 | 52 | 53 | 52 | 53 | 52 | 53 | 52 | 53 | 52 | 53 | 52 | 53 | 52 | 53 | 52 | 53 | 52 | 53 | 52 | 53 | 52 | 53 | 52 | 53 | 52 | 53 | 52 | 53 | 52 | 53 | 52 | 53 | 52 | 53 | 52 | 53 | 52 | 53 | 52 | 53 |
| 53 | 53 | 53 | 53 | 53 | 53 | 53 | 53 | 53 | 53 | 53 | 53 | 53 | 53 | 53 | 53 | 53 | 53 | 53 | 53 | 53 | 53 | 53 | 53 | 53 | 53 | 53 | 53 | 53 | 53 | 53 | 53 | 53 | 53 | 53 | 53 | 53 | 53 | 53 | 53 | 53 | 53 | 53 | 53 | 53 | 53 | 53 | 53 | 53 | 53 | 53 | 53 | 53 | 53 | 53 | 53 | 53 | 53 | 53 | 53 | 53 | 53 | 53 | 53 |
| 54 | 54 | 54 | 54 | 54 | 54 | 54 | 54 | 54 | 54 | 54 | 54 | 54 | 54 | 54 | 54 | 54 | 54 | 54 | 54 | 54 | 54 | 54 | 54 | 54 | 54 | 54 | 54 | 54 | 54 | 54 | 54 | 54 | 54 | 54 | 54 | 54 | 54 | 54 | 54 | 54 | 54 | 54 | 54 | 54 | 54 | 54 | 54 | 54 | 54 | 54 | 54 | 54 | 54 | 54 | 54 | 54 | 54 | 54 | 54 | 54 | 54 | 54 | 54 |
| 55 | 55 | 55 | 55 | 55 | 55 | 55 | 55 | 55 | 55 | 55 | 55 | 55 | 55 | 55 | 55 | 55 | 55 | 55 | 55 | 55 | 55 | 55 | 55 | 55 | 55 | 55 | 55 | 55 | 55 | 55 | 55 | 55 | 55 | 55 | 55 | 55 | 55 | 55 | 55 | 55 | 55 | 55 | 55 | 55 | 55 | 55 | 55 | 55 | 55 | 55 | 55 | 55 | 55 | 55 | 55 | 55 | 55 | 55 | 55 | 55 | 55 | 55 | 55 |
| 56 | 57 | 58 | 58 | 56 | 57 | 58 | 58 | 56 | 57 | 58 | 58 | 56 | 57 | 58 | 58 | 56 | 57 | 58 | 58 | 56 | 57 | 58 | 58 | 56 | 57 | 58 | 58 | 56 | 57 | 58 | 58 | 56 | 57 | 58 | 58 | 56 | 57 | 58 | 58 | 56 | 57 | 58 | 58 | 56 | 57 | 58 | 58 | 56 | 57 | 58 | 58 | 56 | 57 | 58 | 58 | 56 | 57 | 58 | 58 | 56 | 57 | 58 | 58 |
| 57 | 57 | 59 | 59 | 57 | 57 | 59 | 59 | 57 | 57 | 59 | 59 | 57 | 57 | 59 | 59 | 57 | 57 | 59 | 59 | 57 | 57 | 59 | 59 | 57 | 57 | 59 | 59 | 57 | 57 | 59 | 59 | 57 | 57 | 59 | 59 | 57 | 57 | 59 | 59 | 57 | 57 | 59 | 59 | 57 | 57 | 59 | 59 | 57 | 57 | 59 | 59 | 57 | 57 | 59 | 59 | 57 | 57 | 59 | 59 | 57 | 57 | 59 | 59 |
| 58 | 58 | 58 | 58 | 58 | 58 | 58 | 58 | 58 | 58 | 58 | 58 | 58 | 58 | 58 | 58 | 58 | 58 | 58 | 58 | 58 | 58 | 58 | 58 | 58 | 58 | 58 | 58 | 58 | 58 | 58 | 58 | 58 | 58 | 58 | 58 | 58 | 58 | 58 | 58 | 58 | 58 | 58 | 58 | 58 | 58 | 58 | 58 | 58 | 58 | 58 | 58 | 58 | 58 | 58 | 58 | 58 | 58 | 58 | 58 | 58 | 58 | 58 | 58 |
| 59 | 59 | 59 | 59 | 59 | 59 | 59 | 59 | 59 | 59 | 59 | 59 | 59 | 59 | 59 | 59 | 59 | 59 | 59 | 59 | 59 | 59 | 59 | 59 | 59 | 59 | 59 | 59 | 59 | 59 | 59 | 59 | 59 | 59 | 59 | 59 | 59 | 59 | 59 | 59 | 59 | 59 | 59 | 59 | 59 | 59 | 59 | 59 | 59 | 59 | 59 | 59 | 59 | 59 | 59 | 59 | 59 | 59 | 59 | 59 | 59 | 59 | 59 | 59 |
| 60 | 61 | 60 | 61 | 60 | 61 | 60 | 61 | 60 | 61 | 60 | 61 | 60 | 61 | 60 | 61 | 60 | 61 | 60 | 61 | 60 | 61 | 60 | 61 | 60 | 61 | 60 | 61 | 60 | 61 | 60 | 61 | 60 | 61 | 60 | 61 | 60 | 61 | 60 | 61 | 60 | 61 | 60 | 61 | 60 | 61 | 60 | 61 | 60 | 61 | 60 | 61 | 60 | 61 | 60 | 61 | 60 | 61 | 60 | 61 | 60 | 61 | 60 | 61 |
| 61 | 61 | 61 | 61 | 61 | 61 | 61 | 61 | 61 | 61 | 61 | 61 | 61 | 61 | 61 | 61 | 61 | 61 | 61 | 61 | 61 | 61 | 61 | 61 | 61 | 61 | 61 | 61 | 61 | 61 | 61 | 61 | 61 | 61 | 61 | 61 | 61 | 61 | 61 | 61 | 61 | 61 | 61 | 61 | 61 | 61 | 61 | 61 | 61 | 61 | 61 | 61 | 61 | 61 | 61 | 61 | 61 | 61 | 61 | 61 | 61 | 61 | 61 | 61 |
| 62 | 62 | 62 | 62 | 62 | 62 | 62 | 62 | 62 | 62 | 62 | 62 | 62 | 62 | 62 | 62 | 62 | 62 | 62 | 62 | 62 | 62 | 62 | 62 | 62 | 62 | 62 | 62 | 62 | 62 | 62 | 62 | 62 | 62 | 62 | 62 | 62 | 62 | 62 | 62 | 62 | 62 | 62 | 62 | 62 | 62 | 62 | 62 | 62 | 62 | 62 | 62 | 62 | 62 | 62 | 62 | 62 | 62 | 62 | 62 | 62 | 62 | 62 | 62 |
| 63 | 63 | 63 | 63 | 63 | 63 | 63 | 63 | 63 | 63 | 63 | 63 | 63 | 63 | 63 | 63 | 63 | 63 | 63 | 63 | 63 | 63 | 63 | 63 | 63 | 63 | 63 | 63 | 63 | 63 | 63 | 63 | 63 | 63 | 63 | 63 | 63 | 63 | 63 | 63 | 63 | 63 | 63 | 63 | 63 | 63 | 63 | 63 | 63 | 63 | 63 | 63 | 63 | 63 | 63 | 63 | 63 | 63 | 63 | 63 | 63 | 63 | 63 | 63 |

Figure 9: Matrix $M_{L_2}$ of composition $\mathbf{x} \circ \mathbf{y}$ implemented by $L_2$. Symmetric entries are underlined.

| | | | | | | | | | | | | | | | | | | | | | | | | | | | | | | | | | | | | | | | | | | | | | | | | | | | | | | | | | | | | | | | |
|---|---|---|---|---|---|---|---|---|---|---|---|---|---|---|---|---|---|---|---|---|---|---|---|---|---|---|---|---|---|---|---|---|---|---|---|---|---|---|---|---|---|---|---|---|---|---|---|---|---|---|---|---|---|---|---|---|---|---|---|---|---|---|---|
| 0 | 2 | 4 | 6 | 8 | 10 | 12 | 14 | 16 | 18 | 20 | 22 | 24 | 26 | 28 | 30 | 32 | 34 | 36 | 38 | 40 | 42 | 44 | 46 | 48 | 50 | 52 | 54 | 56 | 58 | 60 | 62 | 1 | 3 | 5 | 7 | 9 | 11 | 13 | 15 | 17 | 19 | 21 | 23 | 25 | 27 | 29 | 31 | 33 | 35 | 37 | 39 | 41 | 43 | 45 | 47 | 49 | 51 | 53 | 55 | 57 | 59 | 61 | 63 |
| 1 | 3 | 5 | 7 | 9 | 11 | 13 | 15 | 17 | 19 | 21 | 23 | 25 | 27 | 29 | 31 | 33 | 35 | 37 | 39 | 41 | 43 | 45 | 47 | 49 | 51 | 53 | 55 | 57 | 59 | 61 | 63 | 33 | 35 | 37 | 39 | 41 | 43 | 45 | 47 | 49 | 51 | 53 | 55 | 57 | 59 | 61 | 63 | 49 | 51 | 53 | 55 | 57 | 59 | 61 | 63 | 57 | 59 | 61 | 63 | 61 | 63 | 63 | 63 |
| 2 | 3 | 6 | 7 | 10 | 11 | 14 | 15 | 18 | 19 | 22 | 23 | 26 | 27 | 30 | 31 | 34 | 35 | 38 | 39 | 42 | 43 | 46 | 47 | 50 | 51 | 54 | 55 | 58 | 59 | 62 | 63 | 3 | 3 | 7 | 7 | 11 | 11 | 15 | 15 | 19 | 19 | 23 | 23 | 27 | 27 | 31 | 31 | 35 | 35 | 39 | 39 | 43 | 43 | 47 | 47 | 51 | 51 | 55 | 55 | 59 | 59 | 63 | 63 |
| 3 | 3 | 7 | 7 | 11 | 11 | 15 | 15 | 19 | 19 | 23 | 23 | 27 | 27 | 31 | 31 | 35 | 35 | 39 | 39 | 43 | 43 | 47 | 47 | 51 | 51 | 55 | 55 | 59 | 59 | 63 | 63 | 35 | 35 | 39 | 39 | 43 | 43 | 47 | 47 | 51 | 51 | 55 | 55 | 59 | 59 | 63 | 63 | 51 | 51 | 55 | 55 | 59 | 59 | 63 | 63 | 59 | 59 | 63 | 63 | 63 | 63 | 63 | 63 |
| 4 | 6 | 6 | 7 | 12 | 14 | 14 | 15 | 20 | 22 | 22 | 23 | 28 | 30 | 30 | 31 | 36 | 38 | 38 | 39 | 44 | 46 | 46 | 47 | 52 | 54 | 54 | 55 | 60 | 62 | 62 | 63 | 5 | 7 | 7 | 7 | 13 | 15 | 15 | 15 | 21 | 23 | 23 | 23 | 29 | 31 | 31 | 31 | 37 | 39 | 39 | 39 | 45 | 47 | 47 | 47 | 53 | 55 | 55 | 55 | 61 | 63 | 63 | 63 |
| 5 | 7 | 7 | 7 | 13 | 15 | 15 | 15 | 21 | 23 | 23 | 23 | 29 | 31 | 31 | 31 | 37 | 39 | 39 | 39 | 45 | 47 | 47 | 47 | 53 | 55 | 55 | 55 | 61 | 63 | 63 | 63 | 37 | 39 | 39 | 39 | 45 | 47 | 47 | 47 | 53 | 55 | 55 | 55 | 61 | 63 | 63 | 63 | 53 | 55 | 55 | 55 | 61 | 63 | 63 | 63 | 61 | 63 | 63 | 63 | 61 | 63 | 63 | 63 |
| 6 | 7 | 6 | 7 | 14 | 15 | 14 | 15 | 22 | 23 | 22 | 23 | 30 | 31 | 30 | 31 | 38 | 39 | 38 | 39 | 46 | 47 | 46 | 47 | 54 | 55 | 54 | 55 | 62 | 63 | 62 | 63 | 7 | 7 | 7 | 7 | 15 | 15 | 15 | 15 | 23 | 23 | 23 | 23 | 31 | 31 | 31 | 31 | 39 | 39 | 39 | 39 | 47 | 47 | 47 | 47 | 55 | 55 | 55 | 55 | 63 | 63 | 63 | 63 |
| 7 | 7 | 7 | 7 | 15 | 15 | 15 | 15 | 23 | 23 | 23 | 23 | 31 | 31 | 31 | 31 | 39 | 39 | 39 | 39 | 47 | 47 | 47 | 47 | 55 | 55 | 55 | 55 | 63 | 63 | 63 | 63 | 39 | 39 | 39 | 39 | 47 | 47 | 47 | 47 | 55 | 55 | 55 | 55 | 63 | 63 | 63 | 63 | 55 | 55 | 55 | 55 | 63 | 63 | 63 | 63 | 63 | 63 | 63 | 63 | 63 | 63 | 63 | 63 |
| 8 | 10 | 12 | 14 | 12 | 14 | 14 | 15 | 24 | 26 | 28 | 30 | 28 | 30 | 30 | 31 | 40 | 42 | 44 | 46 | 44 | 46 | 46 | 47 | 56 | 58 | 60 | 62 | 60 | 62 | 62 | 63 | 9 | 11 | 13 | 15 | 13 | 15 | 15 | 15 | 25 | 27 | 29 | 31 | 29 | 31 | 31 | 31 | 41 | 43 | 45 | 47 | 45 | 47 | 47 | 47 | 57 | 59 | 61 | 63 | 61 | 63 | 63 | 63 |
| 9 | 11 | 13 | 15 | 13 | 15 | 15 | 15 | 25 | 27 | 29 | 31 | 29 | 31 | 31 | 31 | 41 | 43 | 45 | 47 | 45 | 47 | 47 | 47 | 57 | 59 | 61 | 63 | 61 | 63 | 63 | 63 | 41 | 43 | 45 | 47 | 45 | 47 | 47 | 47 | 57 | 59 | 61 | 63 | 61 | 63 | 63 | 63 | 57 | 59 | 61 | 63 | 61 | 63 | 63 | 63 | 57 | 59 | 61 | 63 | 61 | 63 | 63 | 63 |
| 10 | 11 | 14 | 15 | 14 | 15 | 14 | 15 | 26 | 27 | 30 | 31 | 30 | 31 | 30 | 31 | 42 | 43 | 46 | 47 | 46 | 47 | 46 | 47 | 58 | 59 | 62 | 63 | 62 | 63 | 62 | 63 | 11 | 11 | 15 | 15 | 15 | 15 | 15 | 15 | 27 | 27 | 31 | 31 | 31 | 31 | 31 | 31 | 43 | 43 | 47 | 47 | 47 | 47 | 47 | 47 | 59 | 59 | 63 | 63 | 63 | 63 | 63 | 63 |
| 11 | 11 | 15 | 15 | 15 | 15 | 15 | 15 | 27 | 27 | 31 | 31 | 31 | 31 | 31 | 31 | 43 | 43 | 47 | 47 | 47 | 47 | 47 | 47 | 59 | 59 | 63 | 63 | 63 | 63 | 63 | 63 | 43 | 43 | 47 | 47 | 47 | 47 | 47 | 47 | 59 | 59 | 63 | 63 | 63 | 63 | 63 | 63 | 59 | 59 | 63 | 63 | 63 | 63 | 63 | 63 | 59 | 59 | 63 | 63 | 63 | 63 | 63 | 63 |
| 12 | 14 | 14 | 15 | 12 | 14 | 14 | 15 | 28 | 30 | 30 | 31 | 28 | 30 | 30 | 31 | 44 | 46 | 46 | 47 | 44 | 46 | 46 | 47 | 60 | 62 | 62 | 63 | 60 | 62 | 62 | 63 | 13 | 15 | 15 | 15 | 13 | 15 | 15 | 15 | 29 | 31 | 31 | 31 | 29 | 31 | 31 | 31 | 45 | 47 | 47 | 47 | 45 | 47 | 47 | 47 | 61 | 63 | 63 | 63 | 61 | 63 | 63 | 63 |
| 13 | 15 | 15 | 15 | 13 | 15 | 15 | 15 | 29 | 31 | 31 | 31 | 29 | 31 | 31 | 31 | 45 | 47 | 47 | 47 | 45 | 47 | 47 | 47 | 61 | 63 | 63 | 63 | 61 | 63 | 63 | 63 | 45 | 47 | 47 | 47 | 45 | 47 | 47 | 47 | 61 | 63 | 63 | 63 | 61 | 63 | 63 | 63 | 61 | 63 | 63 | 63 | 61 | 63 | 63 | 63 | 61 | 63 | 63 | 63 | 61 | 63 | 63 | 63 |
| 14 | 15 | 14 | 15 | 14 | 15 | 14 | 15 | 30 | 31 | 30 | 31 | 30 | 31 | 30 | 31 | 46 | 47 | 46 | 47 | 46 | 47 | 46 | 47 | 62 | 63 | 62 | 63 | 62 | 63 | 62 | 63 | 15 | 15 | 15 | 15 | 15 | 15 | 15 | 15 | 31 | 31 | 31 | 31 | 31 | 31 | 31 | 31 | 47 | 47 | 47 | 47 | 47 | 47 | 47 | 47 | 63 | 63 | 63 | 63 | 63 | 63 | 63 | 63 |
| 15 | 15 | 15 | 15 | 15 | 15 | 15 | 15 | 31 | 31 | 31 | 31 | 31 | 31 | 31 | 31 | 47 | 47 | 47 | 47 | 47 | 47 | 47 | 47 | 63 | 63 | 63 | 63 | 63 | 63 | 63 | 63 | 47 | 47 | 47 | 47 | 47 | 47 | 47 | 47 | 63 | 63 | 63 | 63 | 63 | 63 | 63 | 63 | 63 | 63 | 63 | 63 | 63 | 63 | 63 | 63 | 63 | 63 | 63 | 63 | 63 | 63 | 63 | 63 |
| 16 | 18 | 20 | 22 | 24 | 26 | 28 | 30 | 24 | 26 | 28 | 30 | 28 | 30 | 30 | 31 | 48 | 50 | 52 | 54 | 56 | 58 | 60 | 62 | 56 | 58 | 60 | 62 | 60 | 62 | 62 | 63 | 17 | 19 | 21 | 23 | 25 | 27 | 29 | 31 | 25 | 27 | 29 | 31 | 29 | 31 | 31 | 31 | 49 | 51 | 53 | 55 | 57 | 59 | 61 | 63 | 57 | 59 | 61 | 63 | 61 | 63 | 63 | 63 |
| 17 | 19 | 21 | 23 | 25 | 27 | 29 | 31 | 25 | 27 | 29 | 31 | 29 | 31 | 31 | 31 | 49 | 51 | 53 | 55 | 57 | 59 | 61 | 63 | 57 | 59 | 61 | 63 | 61 | 63 | 63 | 63 | 49 | 51 | 53 | 55 | 57 | 59 | 61 | 63 | 57 | 59 | 61 | 63 | 61 | 63 | 63 | 63 | 49 | 51 | 53 | 55 | 57 | 59 | 61 | 63 | 57 | 59 | 61 | 63 | 61 | 63 | 63 | 63 |
| 18 | 19 | 22 | 23 | 26 | 27 | 30 | 31 | 26 | 27 | 30 | 31 | 30 | 31 | 30 | 31 | 50 | 51 | 54 | 55 | 58 | 59 | 62 | 63 | 58 | 59 | 62 | 63 | 62 | 63 | 62 | 63 | 19 | 19 | 23 | 23 | 27 | 27 | 31 | 31 | 27 | 27 | 31 | 31 | 31 | 31 | 31 | 31 | 51 | 51 | 55 | 55 | 59 | 59 | 63 | 63 | 59 | 59 | 63 | 63 | 63 | 63 | 63 | 63 |
| 19 | 19 | 23 | 23 | 27 | 27 | 31 | 31 | 27 | 27 | 31 | 31 | 31 | 31 | 31 | 31 | 51 | 51 | 55 | 55 | 59 | 59 | 63 | 63 | 59 | 59 | 63 | 63 | 63 | 63 | 63 | 63 | 51 | 51 | 55 | 55 | 59 | 59 | 63 | 63 | 59 | 59 | 63 | 63 | 63 | 63 | 63 | 63 | 51 | 51 | 55 | 55 | 59 | 59 | 63 | 63 | 59 | 59 | 63 | 63 | 63 | 63 | 63 | 63 |
| 20 | 22 | 22 | 23 | 28 | 30 | 30 | 31 | 28 | 30 | 30 | 31 | 28 | 30 | 30 | 31 | 52 | 54 | 54 | 55 | 60 | 62 | 62 | 63 | 60 | 62 | 62 | 63 | 60 | 62 | 62 | 63 | 21 | 23 | 23 | 23 | 29 | 31 | 31 | 31 | 29 | 31 | 31 | 31 | 29 | 31 | 31 | 31 | 53 | 55 | 55 | 55 | 61 | 63 | 63 | 63 | 61 | 63 | 63 | 63 | 61 | 63 | 63 | 63 |
| 21 | 23 | 23 | 23 | 29 | 31 | 31 | 31 | 29 | 31 | 31 | 31 | 29 | 31 | 31 | 31 | 53 | 55 | 55 | 55 | 61 | 63 | 63 | 63 | 61 | 63 | 63 | 63 | 61 | 63 | 63 | 63 | 53 | 55 | 55 | 55 | 61 | 63 | 63 | 63 | 61 | 63 | 63 | 63 | 61 | 63 | 63 | 63 | 53 | 55 | 55 | 55 | 61 | 63 | 63 | 63 | 61 | 63 | 63 | 63 | 61 | 63 | 63 | 63 |
| 22 | 23 | 22 | 23 | 30 | 31 | 30 | 31 | 30 | 31 | 30 | 31 | 30 | 31 | 30 | 31 | 54 | 55 | 54 | 55 | 62 | 63 | 62 | 63 | 62 | 63 | 62 | 63 | 62 | 63 | 62 | 63 | 23 | 23 | 23 | 23 | 31 | 31 | 31 | 31 | 31 | 31 | 31 | 31 | 31 | 31 | 31 | 31 | 55 | 55 | 55 | 55 | 63 | 63 | 63 | 63 | 63 | 63 | 63 | 63 | 63 | 63 | 63 | 63 |
| 23 | 23 | 23 | 23 | 31 | 31 | 31 | 31 | 31 | 31 | 31 | 31 | 31 | 31 | 31 | 31 | 55 | 55 | 55 | 55 | 63 | 63 | 63 | 63 | 63 | 63 | 63 | 63 | 63 | 63 | 63 | 63 | 55 | 55 | 55 | 55 | 63 | 63 | 63 | 63 | 63 | 63 | 63 | 63 | 63 | 63 | 63 | 63 | 55 | 55 | 55 | 55 | 63 | 63 | 63 | 63 | 63 | 63 | 63 | 63 | 63 | 63 | 63 | 63 |
| 24 | 26 | 28 | 30 | 28 | 30 | 30 | 31 | 24 | 26 | 28 | 30 | 28 | 30 | 30 | 31 | 56 | 58 | 60 | 62 | 60 | 62 | 62 | 63 | 56 | 58 | 60 | 62 | 60 | 62 | 62 | 63 | 25 | 27 | 29 | 31 | 29 | 31 | 31 | 31 | 25 | 27 | 29 | 31 | 29 | 31 | 31 | 31 | 57 | 59 | 61 | 63 | 61 | 63 | 63 | 63 | 57 | 59 | 61 | 63 | 61 | 63 | 63 | 63 |
| 25 | 27 | 29 | 31 | 29 | 31 | 31 | 31 | 25 | 27 | 29 | 31 | 29 | 31 | 31 | 31 | 57 | 59 | 61 | 63 | 61 | 63 | 63 | 63 | 57 | 59 | 61 | 63 | 61 | 63 | 63 | 63 | 57 | 59 | 61 | 63 | 61 | 63 | 63 | 63 | 57 | 59 | 61 | 63 | 61 | 63 | 63 | 63 | 57 | 59 | 61 | 63 | 61 | 63 | 63 | 63 | 57 | 59 | 61 | 63 | 61 | 63 | 63 | 63 |
| 26 | 27 | 30 | 31 | 30 | 31 | 30 | 31 | 26 | 27 | 30 | 31 | 30 | 31 | 30 | 31 | 58 | 59 | 62 | 63 | 62 | 63 | 62 | 63 | 58 | 59 | 62 | 63 | 62 | 63 | 62 | 63 | 27 | 27 | 31 | 31 | 31 | 31 | 31 | 31 | 27 | 27 | 31 | 31 | 31 | 31 | 31 | 31 | 59 | 59 | 63 | 63 | 63 | 63 | 63 | 63 | 59 | 59 | 63 | 63 | 63 | 63 | 63 | 63 |
| 27 | 27 | 31 | 31 | 31 | 31 | 31 | 31 | 27 | 27 | 31 | 31 | 31 | 31 | 31 | 31 | 59 | 59 | 63 | 63 | 63 | 63 | 63 | 63 | 59 | 59 | 63 | 63 | 63 | 63 | 63 | 63 | 59 | 59 | 63 | 63 | 63 | 63 | 63 | 63 | 59 | 59 | 63 | 63 | 63 | 63 | 63 | 63 | 59 | 59 | 63 | 63 | 63 | 63 | 63 | 63 | 59 | 59 | 63 | 63 | 63 | 63 | 63 | 63 |
| 28 | 30 | 30 | 31 | 28 | 30 | 30 | 31 | 28 | 30 | 30 | 31 | 28 | 30 | 30 | 31 | 60 | 62 | 62 | 63 | 60 | 62 | 62 | 63 | 60 | 62 | 62 | 63 | 60 | 62 | 62 | 63 | 29 | 31 | 31 | 31 | 29 | 31 | 31 | 31 | 29 | 31 | 31 | 31 | 29 | 31 | 31 | 31 | 61 | 63 | 63 | 63 | 61 | 63 | 63 | 63 | 61 | 63 | 63 | 63 | 61 | 63 | 63 | 63 |
| 29 | 31 | 31 | 31 | 29 | 31 | 31 | 31 | 29 | 31 | 31 | 31 | 29 | 31 | 31 | 31 | 61 | 63 | 63 | 63 | 61 | 63 | 63 | 63 | 61 | 63 | 63 | 63 | 61 | 63 | 63 | 63 | 61 | 63 | 63 | 63 | 61 | 63 | 63 | 63 | 61 | 63 | 63 | 63 | 61 | 63 | 63 | 63 | 61 | 63 | 63 | 63 | 61 | 63 | 63 | 63 | 61 | 63 | 63 | 63 | 61 | 63 | 63 | 63 |
| 30 | 31 | 30 | 31 | 30 | 31 | 30 | 31 | 30 | 31 | 30 | 31 | 30 | 31 | 30 | 31 | 62 | 63 | 62 | 63 | 62 | 63 | 62 | 63 | 62 | 63 | 62 | 63 | 62 | 63 | 62 | 63 | 31 | 31 | 31 | 31 | 31 | 31 | 31 | 31 | 31 | 31 | 31 | 31 | 31 | 31 | 31 | 31 | 63 | 63 | 63 | 63 | 63 | 63 | 63 | 63 | 63 | 63 | 63 | 63 | 63 | 63 | 63 | 63 |
| 31 | 31 | 31 | 31 | 31 | 31 | 31 | 31 | 31 | 31 | 31 | 31 | 31 | 31 | 31 | 31 | 63 | 63 | 63 | 63 | 63 | 63 | 63 | 63 | 63 | 63 | 63 | 63 | 63 | 63 | 63 | 63 | 63 | 63 | 63 | 63 | 63 | 63 | 63 | 63 | 63 | 63 | 63 | 63 | 63 | 63 | 63 | 63 | 63 | 63 | 63 | 63 | 63 | 63 | 63 | 63 | 63 | 63 | 63 | 63 | 63 | 63 | 63 | 63 |
| 32 | 34 | 36 | 38 | 40 | 42 | 44 | 46 | 48 | 50 | 52 | 54 | 56 | 58 | 60 | 62 | 48 | 50 | 52 | 54 | 56 | 58 | 60 | 62 | 56 | 58 | 60 | 62 | 60 | 62 | 62 | 63 | 33 | 35 | 37 | 39 | 41 | 43 | 45 | 47 | 49 | 51 | 53 | 55 | 57 | 59 | 61 | 63 | 49 | 51 | 53 | 55 | 57 | 59 | 61 | 63 | 57 | 59 | 61 | 63 | 61 | 63 | 63 | 63 |
| 33 | 35 | 37 | 39 | 41 | 43 | 45 | 47 | 49 | 51 | 53 | 55 | 57 | 59 | 61 | 63 | 49 | 51 | 53 | 55 | 57 | 59 | 61 | 63 | 57 | 59 | 61 | 63 | 61 | 63 | 63 | 63 | 33 | 35 | 37 | 39 | 41 | 43 | 45 | 47 | 49 | 51 | 53 | 55 | 57 | 59 | 61 | 63 | 49 | 51 | 53 | 55 | 57 | 59 | 61 | 63 | 57 | 59 | 61 | 63 | 61 | 63 | 63 | 63 |
| 34 | 35 | 38 | 39 | 42 | 43 | 46 | 47 | 50 | 51 | 54 | 55 | 58 | 59 | 62 | 63 | 50 | 51 | 54 | 55 | 58 | 59 | 62 | 63 | 58 | 59 | 62 | 63 | 62 | 63 | 62 | 63 | 35 | 35 | 39 | 39 | 43 | 43 | 47 | 47 | 51 | 51 | 55 | 55 | 59 | 59 | 63 | 63 | 51 | 51 | 55 | 55 | 59 | 59 | 63 | 63 | 59 | 59 | 63 | 63 | 63 | 63 | 63 | 63 |
| 35 | 35 | 39 | 39 | 43 | 43 | 47 | 47 | 51 | 51 | 55 | 55 | 59 | 59 | 63 | 63 | 51 | 51 | 55 | 55 | 59 | 59 | 63 | 63 | 59 | 59 | 63 | 63 | 63 | 63 | 63 | 63 | 35 | 35 | 39 | 39 | 43 | 43 | 47 | 47 | 51 | 51 | 55 | 55 | 59 | 59 | 63 | 63 | 51 | 51 | 55 | 55 | 59 | 59 | 63 | 63 | 59 | 59 | 63 | 63 | 63 | 63 | 63 | 63 |
| 36 | 38 | 38 | 39 | 44 | 46 | 46 | 47 | 52 | 54 | 54 | 55 | 60 | 62 | 62 | 63 | 52 | 54 | 54 | 55 | 60 | 62 | 62 | 63 | 60 | 62 | 62 | 63 | 60 | 62 | 62 | 63 | 37 | 39 | 39 | 39 | 45 | 47 | 47 | 47 | 53 | 55 | 55 | 55 | 61 | 63 | 63 | 63 | 53 | 55 | 55 | 55 | 61 | 63 | 63 | 63 | 61 | 63 | 63 | 63 | 61 | 63 | 63 | 63 |
| 37 | 39 | 39 | 39 | 45 | 47 | 47 | 47 | 53 | 55 | 55 | 55 | 61 | 63 | 63 | 63 | 53 | 55 | 55 | 55 | 61 | 63 | 63 | 63 | 61 | 63 | 63 | 63 | 61 | 63 | 63 | 63 | 37 | 39 | 39 | 39 | 45 | 47 | 47 | 47 | 53 | 55 | 55 | 55 | 61 | 63 | 63 | 63 | 53 | 55 | 55 | 55 | 61 | 63 | 63 | 63 | 61 | 63 | 63 | 63 | 61 | 63 | 63 | 63 |
| 38 | 39 | 38 | 39 | 46 | 47 | 46 | 47 | 54 | 55 | 54 | 55 | 62 | 63 | 62 | 63 | 54 | 55 | 54 | 55 | 62 | 63 | 62 | 63 | 62 | 63 | 62 | 63 | 62 | 63 | 62 | 63 | 39 | 39 | 39 | 39 | 47 | 47 | 47 | 47 | 55 | 55 | 55 | 55 | 63 | 63 | 63 | 63 | 55 | 55 | 55 | 55 | 63 | 63 | 63 | 63 | 63 | 63 | 63 | 63 | 63 | 63 | 63 | 63 |
| 39 | 39 | 39 | 39 | 47 | 47 | 47 | 47 | 55 | 55 | 55 | 55 | 63 | 63 | 63 | 63 | 55 | 55 | 55 | 55 | 63 | 63 | 63 | 63 | 63 | 63 | 63 | 63 | 63 | 63 | 63 | 63 | 39 | 39 | 39 | 39 | 47 | 47 | 47 | 47 | 55 | 55 | 55 | 55 | 63 | 63 | 63 | 63 | 55 | 55 | 55 | 55 | 63 | 63 | 63 | 63 | 63 | 63 | 63 | 63 | 63 | 63 | 63 | 63 |
| 40 | 42 | 44 | 46 | 44 | 46 | 46 | 47 | 56 | 58 | 60 | 62 | 60 | 62 | 62 | 63 | 56 | 58 | 60 | 62 | 60 | 62 | 62 | 63 | 56 | 58 | 60 | 62 | 60 | 62 | 62 | 63 | 41 | 43 | 45 | 47 | 45 | 47 | 47 | 47 | 57 | 59 | 61 | 63 | 61 | 63 | 63 | 63 | 57 | 59 | 61 | 63 | 61 | 63 | 63 | 63 | 57 | 59 | 61 | 63 | 61 | 63 | 63 | 63 |
| 41 | 43 | 45 | 47 | 45 | 47 | 47 | 47 | 57 | 59 | 61 | 63 | 61 | 63 | 63 | 63 | 57 | 59 | 61 | 63 | 61 | 63 | 63 | 63 | 57 | 59 | 61 | 63 | 61 | 63 | 63 | 63 | 41 | 43 | 45 | 47 | 45 | 47 | 47 | 47 | 57 | 59 | 61 | 63 | 61 | 63 | 63 | 63 | 57 | 59 | 61 | 63 | 61 | 63 | 63 | 63 | 57 | 59 | 61 | 63 | 61 | 63 | 63 | 63 |
| 42 | 43 | 46 | 47 | 46 | 47 | 46 | 47 | 58 | 59 | 62 | 63 | 62 | 63 | 62 | 63 | 58 | 59 | 62 | 63 | 62 | 63 | 62 | 63 | 58 | 59 | 62 | 63 | 62 | 63 | 62 | 63 | 43 | 43 | 47 | 47 | 47 | 47 | 47 | 47 | 59 | 59 | 63 | 63 | 63 | 63 | 63 | 63 | 59 | 59 | 63 | 63 | 63 | 63 | 63 | 63 | 59 | 59 | 63 | 63 | 63 | 63 | 63 | 63 |
| 43 | 43 | 47 | 47 | 47 | 47 | 47 | 47 | 59 | 59 | 63 | 63 | 63 | 63 | 63 | 63 | 59 | 59 | 63 | 63 | 63 | 63 | 63 | 63 | 59 | 59 | 63 | 63 | 63 | 63 | 63 | 63 | 43 | 43 | 47 | 47 | 47 | 47 | 47 | 47 | 59 | 59 | 63 | 63 | 63 | 63 | 63 | 63 | 59 | 59 | 63 | 63 | 63 | 63 | 63 | 63 | 59 | 59 | 63 | 63 | 63 | 63 | 63 | 63 |
| 44 | 46 | 46 | 47 | 44 | 46 | 46 | 47 | 60 | 62 | 62 | 63 | 60 | 62 | 62 | 63 | 60 | 62 | 62 | 63 | 60 | 62 | 62 | 63 | 60 | 62 | 62 | 63 | 60 | 62 | 62 | 63 | 45 | 47 | 47 | 47 | 45 | 47 | 47 | 47 | 61 | 63 | 63 | 63 | 61 | 63 | 63 | 63 | 61 | 63 | 63 | 63 | 61 | 63 | 63 | 63 | 61 | 63 | 63 | 63 | 61 | 63 | 63 | 63 |
| 45 | 47 | 47 | 47 | 45 | 47 | 47 | 47 | 61 | 63 | 63 | 63 | 61 | 63 | 63 | 63 | 61 | 63 | 63 | 63 | 61 | 63 | 63 | 63 | 61 | 63 | 63 | 63 | 61 | 63 | 63 | 63 | 45 | 47 | 47 | 47 | 45 | 47 | 47 | 47 | 61 | 63 | 63 | 63 | 61 | 63 | 63 | 63 | 61 | 63 | 63 | 63 | 61 | 63 | 63 | 63 | 61 | 63 | 63 | 63 | 61 | 63 | 63 | 63 |
| 46 | 47 | 46 | 47 | 46 | 47 | 46 | 47 | 62 | 63 | 62 | 63 | 62 | 63 | 62 | 63 | 62 | 63 | 62 | 63 | 62 | 63 | 62 | 63 | 62 | 63 | 62 | 63 | 62 | 63 | 62 | 63 | 47 | 47 | 47 | 47 | 47 | 47 | 47 | 47 | 63 | 63 | 63 | 63 | 63 | 63 | 63 | 63 | 63 | 63 | 63 | 63 | 63 | 63 | 63 | 63 | 63 | 63 | 63 | 63 | 63 | 63 | 63 | 63 |
| 47 | 47 | 47 | 47 | 47 | 47 | 47 | 47 | 63 | 63 | 63 | 63 | 63 | 63 | 63 | 63 | 63 | 63 | 63 | 63 | 63 | 63 | 63 | 63 | 63 | 63 | 63 | 63 | 63 | 63 | 63 | 63 | 47 | 47 | 47 | 47 | 47 | 47 | 47 | 47 | 63 | 63 | 63 | 63 | 63 | 63 | 63 | 63 | 63 | 63 | 63 | 63 | 63 | 63 | 63 | 63 | 63 | 63 | 63 | 63 | 63 | 63 | 63 | 63 |
| 48 | 50 | 52 | 54 | 56 | 58 | 60 | 62 | 56 | 58 | 60 | 62 | 60 | 62 | 62 | 63 | 48 | 50 | 52 | 54 | 56 | 58 | 60 | 62 | 56 | 58 | 60 | 62 | 60 | 62 | 62 | 63 | 49 | 51 | 53 | 55 | 57 | 59 | 61 | 63 | 57 | 59 | 61 | 63 | 61 | 63 | 63 | 63 | 49 | 51 | 53 | 55 | 57 | 59 | 61 | 63 | 57 | 59 | 61 | 63 | 61 | 63 | 63 | 63 |
| 49 | 51 | 53 | 55 | 57 | 59 | 61 | 63 | 57 | 59 | 61 | 63 | 61 | 63 | 63 | 63 | 49 | 51 | 53 | 55 | 57 | 59 | 61 | 63 | 57 | 59 | 61 | 63 | 61 | 63 | 63 | 63 | 49 | 51 | 53 | 55 | 57 | 59 | 61 | 63 | 57 | 59 | 61 | 63 | 61 | 63 | 63 | 63 | 49 | 51 | 53 | 55 | 57 | 59 | 61 | 63 | 57 | 59 | 61 | 63 | 61 | 63 | 63 | 63 |
| 50 | 51 | 54 | 55 | 58 | 59 | 62 | 63 | 58 | 59 | 62 | 63 | 62 | 63 | 62 | 63 | 50 | 51 | 54 | 55 | 58 | 59 | 62 | 63 | 58 | 59 | 62 | 63 | 62 | 63 | 62 | 63 | 51 | 51 | 55 | 55 | 59 | 59 | 63 | 63 | 59 | 59 | 63 | 63 | 63 | 63 | 63 | 63 | 51 | 51 | 55 | 55 | 59 | 59 | 63 | 63 | 59 | 59 | 63 | 63 | 63 | 63 | 63 | 63 |
| 51 | 51 | 55 | 55 | 59 | 59 | 63 | 63 | 59 | 59 | 63 | 63 | 63 | 63 | 63 | 63 | 51 | 51 | 55 | 55 | 59 | 59 | 63 | 63 | 59 | 59 | 63 | 63 | 63 | 63 | 63 | 63 | 51 | 51 | 55 | 55 | 59 | 59 | 63 | 63 | 59 | 59 | 63 | 63 | 63 | 63 | 63 | 63 | 51 | 51 | 55 | 55 | 59 | 59 | 63 | 63 | 59 | 59 | 63 | 63 | 63 | 63 | 63 | 63 |
| 52 | 54 | 54 | 55 | 60 | 62 | 62 | 63 | 60 | 62 | 62 | 63 | 60 | 62 | 62 | 63 | 52 | 54 | 54 | 55 | 60 | 62 | 62 | 63 | 60 | 62 | 62 | 63 | 60 | 62 | 62 | 63 | 53 | 55 | 55 | 55 | 61 | 63 | 63 | 63 | 61 | 63 | 63 | 63 | 61 | 63 | 63 | 63 | 53 | 55 | 55 | 55 | 61 | 63 | 63 | 63 | 61 | 63 | 63 | 63 | 61 | 63 | 63 | 63 |
| 53 | 55 | 55 | 55 | 61 | 63 | 63 | 63 | 61 | 63 | 63 | 63 | 61 | 63 | 63 | 63 | 53 | 55 | 55 | 55 | 61 | 63 | 63 | 63 | 61 | 63 | 63 | 63 | 61 | 63 | 63 | 63 | 53 | 55 | 55 | 55 | 61 | 63 | 63 | 63 | 61 | 63 | 63 | 63 | 61 | 63 | 63 | 63 | 53 | 55 | 55 | 55 | 61 | 63 | 63 | 63 | 61 | 63 | 63 | 63 | 61 | 63 | 63 | 63 |
| 54 | 55 | 54 | 55 | 62 | 63 | 62 | 63 | 62 | 63 | 62 | 63 | 62 | 63 | 62 | 63 | 54 | 55 | 54 | 55 | 62 | 63 | 62 | 63 | 62 | 63 | 62 | 63 | 62 | 63 | 62 | 63 | 55 | 55 | 55 | 55 | 63 | 63 | 63 | 63 | 63 | 63 | 63 | 63 | 63 | 63 | 63 | 63 | 55 | 55 | 55 | 55 | 63 | 63 | 63 | 63 | 63 | 63 | 63 | 63 | 63 | 63 | 63 | 63 |
| 55 | 55 | 55 | 55 | 63 | 63 | 63 | 63 | 63 | 63 | 63 | 63 | 63 | 63 | 63 | 63 | 55 | 55 | 55 | 55 | 63 | 63 | 63 | 63 | 63 | 63 | 63 | 63 | 63 | 63 | 63 | 63 | 55 | 55 | 55 | 55 | 63 | 63 | 63 | 63 | 63 | 63 | 63 | 63 | 63 | 63 | 63 | 63 | 55 | 55 | 55 | 55 | 63 | 63 | 63 | 63 | 63 | 63 | 63 | 63 | 63 | 63 | 63 | 63 |
| 56 | 58 | 60 | 62 | 60 | 62 | 62 | 63 | 56 | 58 | 60 | 62 | 60 | 62 | 62 | 63 | 56 | 58 | 60 | 62 | 60 | 62 | 62 | 63 | 56 | 58 | 60 | 62 | 60 | 62 | 62 | 63 | 57 | 59 | 61 | 63 | 61 | 63 | 63 | 63 | 57 | 59 | 61 | 63 | 61 | 63 | 63 | 63 | 57 | 59 | 61 | 63 | 61 | 63 | 63 | 63 | 57 | 59 | 61 | 63 | 61 | 63 | 63 | 63 |
| 57 | 59 | 61 | 63 | 61 | 63 | 63 | 63 | 57 | 59 | 61 | 63 | 61 | 63 | 63 | 63 | 57 | 59 | 61 | 63 | 61 | 63 | 63 | 63 | 57 | 59 | 61 | 63 | 61 | 63 | 63 | 63 | 57 | 59 | 61 | 63 | 61 | 63 | 63 | 63 | 57 | 59 | 61 | 63 | 61 | 63 | 63 | 63 | 57 | 59 | 61 | 63 | 61 | 63 | 63 | 63 | 57 | 59 | 61 | 63 | 61 | 63 | 63 | 63 |
| 58 | 59 | 62 | 63 | 62 | 63 | 62 | 63 | 58 | 59 | 62 | 63 | 62 | 63 | 62 | 63 | 58 | 59 | 62 | 63 | 62 | 63 | 62 | 63 | 58 | 59 | 62 | 63 | 62 | 63 | 62 | 63 | 59 | 59 | 63 | 63 | 63 | 63 | 63 | 63 | 59 | 59 | 63 | 63 | 63 | 63 | 63 | 63 | 59 | 59 | 63 | 63 | 63 | 63 | 63 | 63 | 59 | 59 | 63 | 63 | 63 | 63 | 63 | 63 |
| 59 | 59 | 63 | 63 | 63 | 63 | 63 | 63 | 59 | 59 | 63 | 63 | 63 | 63 | 63 | 63 | 59 | 59 | 63 | 63 | 63 | 63 | 63 | 63 | 59 | 59 | 63 | 63 | 63 | 63 | 63 | 63 | 59 | 59 | 63 | 63 | 63 | 63 | 63 | 63 | 59 | 59 | 63 | 63 | 63 | 63 | 63 | 63 | 59 | 59 | 63 | 63 | 63 | 63 | 63 | 63 | 59 | 59 | 63 | 63 | 63 | 63 | 63 | 63 |
| 60 | 62 | 62 | 63 | 60 | 62 | 62 | 63 | 60 | 62 | 62 | 63 | 60 | 62 | 62 | 63 | 60 | 62 | 62 | 63 | 60 | 62 | 62 | 63 | 60 | 62 | 62 | 63 | 60 | 62 | 62 | 63 | 61 | 63 | 63 | 63 | 61 | 63 | 63 | 63 | 61 | 63 | 63 | 63 | 61 | 63 | 63 | 63 | 61 | 63 | 63 | 63 | 61 | 63 | 63 | 63 | 61 | 63 | 63 | 63 | 61 | 63 | 63 | 63 |
| 61 | 63 | 63 | 63 | 61 | 63 | 63 | 63 | 61 | 63 | 63 | 63 | 61 | 63 | 63 | 63 | 61 | 63 | 63 | 63 | 61 | 63 | 63 | 63 | 61 | 63 | 63 | 63 | 61 | 63 | 63 | 63 | 61 | 63 | 63 | 63 | 61 | 63 | 63 | 63 | 61 | 63 | 63 | 63 | 61 | 63 | 63 | 63 | 61 | 63 | 63 | 63 | 61 | 63 | 63 | 63 | 61 | 63 | 63 | 63 | 61 | 63 | 63 | 63 |
| 62 | 63 | 62 | 63 | 62 | 63 | 62 | 63 | 62 | 63 | 62 | 63 | 62 | 63 | 62 | 63 | 62 | 63 | 62 | 63 | 62 | 63 | 62 | 63 | 62 | 63 | 62 | 63 | 62 | 63 | 62 | 63 | 63 | 63 | 63 | 63 | 63 | 63 | 63 | 63 | 63 | 63 | 63 | 63 | 63 | 63 | 63 | 63 | 63 | 63 | 63 | 63 | 63 | 63 | 63 | 63 | 63 | 63 | 63 | 63 | 63 | 63 | 63 | 63 |
| 63 | 63 | 63 | 63 | 63 | 63 | 63 | 63 | 63 | 63 | 63 | 63 | 63 | 63 | 63 | 63 | 63 | 63 | 63 | 63 | 63 | 63 | 63 | 63 | 63 | 63 | 63 | 63 | 63 | 63 | 63 | 63 | 63 | 63 | 63 | 63 | 63 | 63 | 63 | 63 | 63 | 63 | 63 | 63 | 63 | 63 | 63 | 63 | 63 | 63 | 63 | 63 | 63 | 63 | 63 | 63 | 63 | 63 | 63 | 63 | 63 | 63 | 63 | 63 |

Figure 10: Matrix $M_{L_3}$ of composition $\mathbf{x} \circ \mathbf{y}$ implemented by $L_3$. Symmetric entries are underlined.

| | | | | | | | | | | | | | | | | | | | | | | | | | | | | | | | | | | | | | | | | | | | | | | | | | | | | | | | | | | | | | |
|---|---|---|---|---|---|---|---|---|---|---|---|---|---|---|---|---|---|---|---|---|---|---|---|---|---|---|---|---|---|---|---|---|---|---|---|---|---|---|---|---|---|---|---|---|---|---|---|---|---|---|---|---|---|---|---|---|---|---|---|---|---|---|---|
| 0 | 3 | 6 | 6 | 12 | 15 | 12 | 15 | 24 | 27 | 30 | 30 | 24 | 27 | 30 | 30 | 48 | 51 | 54 | 54 | 60 | 63 | 60 | 63 | 48 | 51 | 54 | 54 | 60 | 63 | 60 | 63 | 33 | 35 | 39 | 39 | 45 | 47 | 45 | 47 | 57 | 59 | 63 | 63 | 57 | 59 | 63 | 63 | 33 | 35 | 39 | 39 | 45 | 47 | 45 | 47 | 57 | 59 | 63 | 63 | 57 | 59 | 63 | 63 |
| 1 | 3 | 7 | 7 | 13 | 15 | 13 | 15 | 25 | 27 | 31 | 31 | 25 | 27 | 31 | 31 | 49 | 51 | 55 | 55 | 61 | 63 | 61 | 63 | 49 | 51 | 55 | 55 | 61 | 63 | 61 | 63 | 1 | 3 | 7 | 7 | 13 | 15 | 13 | 15 | 25 | 27 | 31 | 31 | 25 | 27 | 31 | 31 | 49 | 51 | 55 | 55 | 61 | 63 | 61 | 63 | 49 | 51 | 55 | 55 | 61 | 63 | 61 | 63 |
| 2 | 2 | 6 | 6 | 14 | 14 | 14 | 14 | 26 | 26 | 30 | 30 | 26 | 26 | 30 | 30 | 50 | 50 | 54 | 54 | 62 | 62 | 62 | 62 | 50 | 50 | 54 | 54 | 62 | 62 | 62 | 62 | 35 | 35 | 39 | 39 | 47 | 47 | 47 | 47 | 59 | 59 | 63 | 63 | 59 | 59 | 63 | 63 | 35 | 35 | 39 | 39 | 47 | 47 | 47 | 47 | 59 | 59 | 63 | 63 | 59 | 59 | 63 | 63 |
| 3 | 3 | 7 | 7 | 15 | 15 | 15 | 15 | 27 | 27 | 31 | 31 | 27 | 27 | 31 | 31 | 51 | 51 | 55 | 55 | 63 | 63 | 63 | 63 | 51 | 51 | 55 | 55 | 63 | 63 | 63 | 63 | 3 | 3 | 7 | 7 | 15 | 15 | 15 | 15 | 27 | 27 | 31 | 31 | 27 | 27 | 31 | 31 | 51 | 51 | 55 | 55 | 63 | 63 | 63 | 63 | 51 | 51 | 55 | 55 | 63 | 63 | 63 | 63 |
| 4 | 7 | 4 | 7 | 12 | 15 | 12 | 15 | 28 | 31 | 28 | 31 | 28 | 31 | 28 | 31 | 52 | 55 | 52 | 55 | 60 | 63 | 60 | 63 | 52 | 55 | 52 | 55 | 60 | 63 | 60 | 63 | 37 | 39 | 37 | 39 | 45 | 47 | 45 | 47 | 61 | 63 | 61 | 63 | 61 | 63 | 61 | 63 | 37 | 39 | 37 | 39 | 45 | 47 | 45 | 47 | 61 | 63 | 61 | 63 | 61 | 63 | 61 | 63 |
| 5 | 7 | 5 | 7 | 13 | 15 | 13 | 15 | 29 | 31 | 29 | 31 | 29 | 31 | 29 | 31 | 53 | 55 | 53 | 55 | 61 | 63 | 61 | 63 | 53 | 55 | 53 | 55 | 61 | 63 | 61 | 63 | 5 | 7 | 5 | 7 | 13 | 15 | 13 | 15 | 29 | 31 | 29 | 31 | 29 | 31 | 29 | 31 | 53 | 55 | 53 | 55 | 61 | 63 | 61 | 63 | 53 | 55 | 53 | 55 | 61 | 63 | 61 | 63 |
| 6 | 6 | 6 | 6 | 14 | 14 | 14 | 14 | 30 | 30 | 30 | 30 | 30 | 30 | 30 | 30 | 54 | 54 | 54 | 54 | 62 | 62 | 62 | 62 | 54 | 54 | 54 | 54 | 62 | 62 | 62 | 62 | 39 | 39 | 39 | 39 | 47 | 47 | 47 | 47 | 63 | 63 | 63 | 63 | 63 | 63 | 63 | 63 | 39 | 39 | 39 | 39 | 47 | 47 | 47 | 47 | 63 | 63 | 63 | 63 | 63 | 63 | 63 | 63 |
| 7 | 7 | 7 | 7 | 15 | 15 | 15 | 15 | 31 | 31 | 31 | 31 | 31 | 31 | 31 | 31 | 55 | 55 | 55 | 55 | 63 | 63 | 63 | 63 | 55 | 55 | 55 | 55 | 63 | 63 | 63 | 63 | 7 | 7 | 7 | 7 | 15 | 15 | 15 | 15 | 31 | 31 | 31 | 31 | 31 | 31 | 31 | 31 | 55 | 55 | 55 | 55 | 63 | 63 | 63 | 63 | 55 | 55 | 55 | 55 | 63 | 63 | 63 | 63 |
| 8 | 11 | 14 | 14 | 8 | 11 | 14 | 14 | 24 | 27 | 30 | 30 | 24 | 27 | 30 | 30 | 56 | 59 | 62 | 62 | 56 | 59 | 62 | 62 | 56 | 59 | 62 | 62 | 56 | 59 | 62 | 62 | 41 | 43 | 47 | 47 | 41 | 43 | 47 | 47 | 57 | 59 | 63 | 63 | 57 | 59 | 63 | 63 | 41 | 43 | 47 | 47 | 41 | 43 | 47 | 47 | 57 | 59 | 63 | 63 | 57 | 59 | 63 | 63 |
| 9 | 11 | 15 | 15 | 9 | 11 | 15 | 15 | 25 | 27 | 31 | 31 | 25 | 27 | 31 | 31 | 57 | 59 | 63 | 63 | 57 | 59 | 63 | 63 | 57 | 59 | 63 | 63 | 57 | 59 | 63 | 63 | 9 | 11 | 15 | 15 | 9 | 11 | 15 | 15 | 25 | 27 | 31 | 31 | 25 | 27 | 31 | 31 | 57 | 59 | 63 | 63 | 57 | 59 | 63 | 63 | 57 | 59 | 63 | 63 | 57 | 59 | 63 | 63 |
| 10 | 10 | 14 | 14 | 10 | 10 | 14 | 14 | 26 | 26 | 30 | 30 | 26 | 26 | 30 | 30 | 58 | 58 | 62 | 62 | 58 | 58 | 62 | 62 | 58 | 58 | 62 | 62 | 58 | 58 | 62 | 62 | 43 | 43 | 47 | 47 | 43 | 43 | 47 | 47 | 59 | 59 | 63 | 63 | 59 | 59 | 63 | 63 | 43 | 43 | 47 | 47 | 43 | 43 | 47 | 47 | 59 | 59 | 63 | 63 | 59 | 59 | 63 | 63 |
| 11 | 11 | 15 | 15 | 11 | 11 | 15 | 15 | 27 | 27 | 31 | 31 | 27 | 27 | 31 | 31 | 59 | 59 | 63 | 63 | 59 | 59 | 63 | 63 | 59 | 59 | 63 | 63 | 59 | 59 | 63 | 63 | 11 | 11 | 15 | 15 | 11 | 11 | 15 | 15 | 27 | 27 | 31 | 31 | 27 | 27 | 31 | 31 | 59 | 59 | 63 | 63 | 59 | 59 | 63 | 63 | 59 | 59 | 63 | 63 | 59 | 59 | 63 | 63 |
| 12 | 15 | 12 | 15 | 12 | 15 | 12 | 15 | 28 | 31 | 28 | 31 | 28 | 31 | 28 | 31 | 60 | 63 | 60 | 63 | 60 | 63 | 60 | 63 | 60 | 63 | 60 | 63 | 60 | 63 | 60 | 63 | 45 | 47 | 45 | 47 | 45 | 47 | 45 | 47 | 61 | 63 | 61 | 63 | 61 | 63 | 61 | 63 | 45 | 47 | 45 | 47 | 45 | 47 | 45 | 47 | 61 | 63 | 61 | 63 | 61 | 63 | 61 | 63 |
| 13 | 15 | 13 | 15 | 13 | 15 | 13 | 15 | 29 | 31 | 29 | 31 | 29 | 31 | 29 | 31 | 61 | 63 | 61 | 63 | 61 | 63 | 61 | 63 | 61 | 63 | 61 | 63 | 61 | 63 | 61 | 63 | 13 | 15 | 13 | 15 | 13 | 15 | 13 | 15 | 29 | 31 | 29 | 31 | 29 | 31 | 29 | 31 | 61 | 63 | 61 | 63 | 61 | 63 | 61 | 63 | 61 | 63 | 61 | 63 | 61 | 63 | 61 | 63 |
| 14 | 14 | 14 | 14 | 14 | 14 | 14 | 14 | 30 | 30 | 30 | 30 | 30 | 30 | 30 | 30 | 62 | 62 | 62 | 62 | 62 | 62 | 62 | 62 | 62 | 62 | 62 | 62 | 62 | 62 | 62 | 62 | 47 | 47 | 47 | 47 | 47 | 47 | 47 | 47 | 63 | 63 | 63 | 63 | 63 | 63 | 63 | 63 | 47 | 47 | 47 | 47 | 47 | 47 | 47 | 47 | 63 | 63 | 63 | 63 | 63 | 63 | 63 | 63 |
| 15 | 15 | 15 | 15 | 15 | 15 | 15 | 15 | 31 | 31 | 31 | 31 | 31 | 31 | 31 | 31 | 63 | 63 | 63 | 63 | 63 | 63 | 63 | 63 | 63 | 63 | 63 | 63 | 63 | 63 | 63 | 63 | 15 | 15 | 15 | 15 | 15 | 15 | 15 | 15 | 31 | 31 | 31 | 31 | 31 | 31 | 31 | 31 | 63 | 63 | 63 | 63 | 63 | 63 | 63 | 63 | 63 | 63 | 63 | 63 | 63 | 63 | 63 | 63 |
| 16 | 19 | 22 | 22 | 28 | 31 | 28 | 31 | 16 | 19 | 22 | 22 | 28 | 31 | 28 | 31 | 48 | 51 | 54 | 54 | 60 | 63 | 60 | 63 | 48 | 51 | 54 | 54 | 60 | 63 | 60 | 63 | 49 | 51 | 55 | 55 | 61 | 63 | 61 | 63 | 49 | 51 | 55 | 55 | 61 | 63 | 61 | 63 | 49 | 51 | 55 | 55 | 61 | 63 | 61 | 63 | 49 | 51 | 55 | 55 | 61 | 63 | 61 | 63 |
| 17 | 19 | 23 | 23 | 29 | 31 | 29 | 31 | 17 | 19 | 23 | 23 | 29 | 31 | 29 | 31 | 49 | 51 | 55 | 55 | 61 | 63 | 61 | 63 | 49 | 51 | 55 | 55 | 61 | 63 | 61 | 63 | 17 | 19 | 23 | 23 | 29 | 31 | 29 | 31 | 17 | 19 | 23 | 23 | 29 | 31 | 29 | 31 | 49 | 51 | 55 | 55 | 61 | 63 | 61 | 63 | 49 | 51 | 55 | 55 | 61 | 63 | 61 | 63 |
| 18 | 18 | 22 | 22 | 30 | 30 | 30 | 30 | 18 | 18 | 22 | 22 | 30 | 30 | 30 | 30 | 50 | 50 | 54 | 54 | 62 | 62 | 62 | 62 | 50 | 50 | 54 | 54 | 62 | 62 | 62 | 62 | 51 | 51 | 55 | 55 | 63 | 63 | 63 | 63 | 51 | 51 | 55 | 55 | 63 | 63 | 63 | 63 | 51 | 51 | 55 | 55 | 63 | 63 | 63 | 63 | 51 | 51 | 55 | 55 | 63 | 63 | 63 | 63 |
| 19 | 19 | 23 | 23 | 31 | 31 | 31 | 31 | 19 | 19 | 23 | 23 | 31 | 31 | 31 | 31 | 51 | 51 | 55 | 55 | 63 | 63 | 63 | 63 | 51 | 51 | 55 | 55 | 63 | 63 | 63 | 63 | 19 | 19 | 23 | 23 | 31 | 31 | 31 | 31 | 19 | 19 | 23 | 23 | 31 | 31 | 31 | 31 | 51 | 51 | 55 | 55 | 63 | 63 | 63 | 63 | 51 | 51 | 55 | 55 | 63 | 63 | 63 | 63 |
| 20 | 23 | 20 | 23 | 28 | 31 | 28 | 31 | 20 | 23 | 20 | 23 | 28 | 31 | 28 | 31 | 52 | 55 | 52 | 55 | 60 | 63 | 60 | 63 | 52 | 55 | 52 | 55 | 60 | 63 | 60 | 63 | 53 | 55 | 53 | 55 | 61 | 63 | 61 | 63 | 53 | 55 | 53 | 55 | 61 | 63 | 61 | 63 | 53 | 55 | 53 | 55 | 61 | 63 | 61 | 63 | 53 | 55 | 53 | 55 | 61 | 63 | 61 | 63 |
| 21 | 23 | 21 | 23 | 29 | 31 | 29 | 31 | 21 | 23 | 21 | 23 | 29 | 31 | 29 | 31 | 53 | 55 | 53 | 55 | 61 | 63 | 61 | 63 | 53 | 55 | 53 | 55 | 61 | 63 | 61 | 63 | 21 | 23 | 21 | 23 | 29 | 31 | 29 | 31 | 21 | 23 | 21 | 23 | 29 | 31 | 29 | 31 | 53 | 55 | 53 | 55 | 61 | 63 | 61 | 63 | 53 | 55 | 53 | 55 | 61 | 63 | 61 | 63 |
| 22 | 22 | 22 | 22 | 30 | 30 | 30 | 30 | 22 | 22 | 22 | 22 | 30 | 30 | 30 | 30 | 54 | 54 | 54 | 54 | 62 | 62 | 62 | 62 | 54 | 54 | 54 | 54 | 62 | 62 | 62 | 62 | 55 | 55 | 55 | 55 | 63 | 63 | 63 | 63 | 55 | 55 | 55 | 55 | 63 | 63 | 63 | 63 | 55 | 55 | 55 | 55 | 63 | 63 | 63 | 63 | 55 | 55 | 55 | 55 | 63 | 63 | 63 | 63 |
| 23 | 23 | 23 | 23 | 31 | 31 | 31 | 31 | 23 | 23 | 23 | 23 | 31 | 31 | 31 | 31 | 55 | 55 | 55 | 55 | 63 | 63 | 63 | 63 | 55 | 55 | 55 | 55 | 63 | 63 | 63 | 63 | 23 | 23 | 23 | 23 | 31 | 31 | 31 | 31 | 23 | 23 | 23 | 23 | 31 | 31 | 31 | 31 | 55 | 55 | 55 | 55 | 63 | 63 | 63 | 63 | 55 | 55 | 55 | 55 | 63 | 63 | 63 | 63 |
| 24 | 27 | 30 | 30 | 24 | 27 | 30 | 30 | 24 | 27 | 30 | 30 | 24 | 27 | 30 | 30 | 56 | 59 | 62 | 62 | 56 | 59 | 62 | 62 | 56 | 59 | 62 | 62 | 56 | 59 | 62 | 62 | 57 | 59 | 63 | 63 | 57 | 59 | 63 | 63 | 57 | 59 | 63 | 63 | 57 | 59 | 63 | 63 | 57 | 59 | 63 | 63 | 57 | 59 | 63 | 63 | 57 | 59 | 63 | 63 | 57 | 59 | 63 | 63 |
| 25 | 27 | 31 | 31 | 25 | 27 | 31 | 31 | 25 | 27 | 31 | 31 | 25 | 27 | 31 | 31 | 57 | 59 | 63 | 63 | 57 | 59 | 63 | 63 | 57 | 59 | 63 | 63 | 57 | 59 | 63 | 63 | 25 | 27 | 31 | 31 | 25 | 27 | 31 | 31 | 25 | 27 | 31 | 31 | 25 | 27 | 31 | 31 | 57 | 59 | 63 | 63 | 57 | 59 | 63 | 63 | 57 | 59 | 63 | 63 | 57 | 59 | 63 | 63 |
| 26 | 26 | 30 | 30 | 26 | 26 | 30 | 30 | 26 | 26 | 30 | 30 | 26 | 26 | 30 | 30 | 58 | 58 | 62 | 62 | 58 | 58 | 62 | 62 | 58 | 58 | 62 | 62 | 58 | 58 | 62 | 62 | 59 | 59 | 63 | 63 | 59 | 59 | 63 | 63 | 59 | 59 | 63 | 63 | 59 | 59 | 63 | 63 | 59 | 59 | 63 | 63 | 59 | 59 | 63 | 63 | 59 | 59 | 63 | 63 | 59 | 59 | 63 | 63 |
| 27 | 27 | 31 | 31 | 27 | 27 | 31 | 31 | 27 | 27 | 31 | 31 | 27 | 27 | 31 | 31 | 59 | 59 | 63 | 63 | 59 | 59 | 63 | 63 | 59 | 59 | 63 | 63 | 59 | 59 | 63 | 63 | 27 | 27 | 31 | 31 | 27 | 27 | 31 | 31 | 27 | 27 | 31 | 31 | 27 | 27 | 31 | 31 | 59 | 59 | 63 | 63 | 59 | 59 | 63 | 63 | 59 | 59 | 63 | 63 | 59 | 59 | 63 | 63 |
| 28 | 31 | 28 | 31 | 28 | 31 | 28 | 31 | 28 | 31 | 28 | 31 | 28 | 31 | 28 | 31 | 60 | 63 | 60 | 63 | 60 | 63 | 60 | 63 | 60 | 63 | 60 | 63 | 60 | 63 | 60 | 63 | 61 | 63 | 61 | 63 | 61 | 63 | 61 | 63 | 61 | 63 | 61 | 63 | 61 | 63 | 61 | 63 | 61 | 63 | 61 | 63 | 61 | 63 | 61 | 63 | 61 | 63 | 61 | 63 | 61 | 63 | 61 | 63 |
| 29 | 31 | 29 | 31 | 29 | 31 | 29 | 31 | 29 | 31 | 29 | 31 | 29 | 31 | 29 | 31 | 61 | 63 | 61 | 63 | 61 | 63 | 61 | 63 | 61 | 63 | 61 | 63 | 61 | 63 | 61 | 63 | 29 | 31 | 29 | 31 | 29 | 31 | 29 | 31 | 29 | 31 | 29 | 31 | 29 | 31 | 29 | 31 | 61 | 63 | 61 | 63 | 61 | 63 | 61 | 63 | 61 | 63 | 61 | 63 | 61 | 63 | 61 | 63 |
| 30 | 30 | 30 | 30 | 30 | 30 | 30 | 30 | 30 | 30 | 30 | 30 | 30 | 30 | 30 | 30 | 62 | 62 | 62 | 62 | 62 | 62 | 62 | 62 | 62 | 62 | 62 | 62 | 62 | 62 | 62 | 62 | 63 | 63 | 63 | 63 | 63 | 63 | 63 | 63 | 63 | 63 | 63 | 63 | 63 | 63 | 63 | 63 | 63 | 63 | 63 | 63 | 63 | 63 | 63 | 63 | 63 | 63 | 63 | 63 | 63 | 63 | 63 | 63 |
| 31 | 31 | 31 | 31 | 31 | 31 | 31 | 31 | 31 | 31 | 31 | 31 | 31 | 31 | 31 | 31 | 63 | 63 | 63 | 63 | 63 | 63 | 63 | 63 | 63 | 63 | 63 | 63 | 63 | 63 | 63 | 63 | 31 | 31 | 31 | 31 | 31 | 31 | 31 | 31 | 31 | 31 | 31 | 31 | 31 | 31 | 31 | 31 | 63 | 63 | 63 | 63 | 63 | 63 | 63 | 63 | 63 | 63 | 63 | 63 | 63 | 63 | 63 | 63 |
| 32 | 35 | 38 | 38 | 44 | 47 | 44 | 47 | 56 | 59 | 62 | 62 | 56 | 59 | 62 | 62 | 32 | 35 | 38 | 38 | 44 | 47 | 44 | 47 | 56 | 59 | 62 | 62 | 56 | 59 | 62 | 62 | 33 | 35 | 39 | 39 | 45 | 47 | 45 | 47 | 57 | 59 | 63 | 63 | 57 | 59 | 63 | 63 | 33 | 35 | 39 | 39 | 45 | 47 | 45 | 47 | 57 | 59 | 63 | 63 | 57 | 59 | 63 | 63 |
| 33 | 35 | 39 | 39 | 45 | 47 | 45 | 47 | 57 | 59 | 63 | 63 | 57 | 59 | 63 | 63 | 33 | 35 | 39 | 39 | 45 | 47 | 45 | 47 | 57 | 59 | 63 | 63 | 57 | 59 | 63 | 63 | 33 | 35 | 39 | 39 | 45 | 47 | 45 | 47 | 57 | 59 | 63 | 63 | 57 | 59 | 63 | 63 | 33 | 35 | 39 | 39 | 45 | 47 | 45 | 47 | 57 | 59 | 63 | 63 | 57 | 59 | 63 | 63 |
| 34 | 34 | 38 | 38 | 46 | 46 | 46 | 46 | 58 | 58 | 62 | 62 | 58 | 58 | 62 | 62 | 34 | 34 | 38 | 38 | 46 | 46 | 46 | 46 | 58 | 58 | 62 | 62 | 58 | 58 | 62 | 62 | 35 | 35 | 39 | 39 | 47 | 47 | 47 | 47 | 59 | 59 | 63 | 63 | 59 | 59 | 63 | 63 | 35 | 35 | 39 | 39 | 47 | 47 | 47 | 47 | 59 | 59 | 63 | 63 | 59 | 59 | 63 | 63 |
| 35 | 35 | 39 | 39 | 47 | 47 | 47 | 47 | 59 | 59 | 63 | 63 | 59 | 59 | 63 | 63 | 35 | 35 | 39 | 39 | 47 | 47 | 47 | 47 | 59 | 59 | 63 | 63 | 59 | 59 | 63 | 63 | 35 | 35 | 39 | 39 | 47 | 47 | 47 | 47 | 59 | 59 | 63 | 63 | 59 | 59 | 63 | 63 | 35 | 35 | 39 | 39 | 47 | 47 | 47 | 47 | 59 | 59 | 63 | 63 | 59 | 59 | 63 | 63 |
| 36 | 39 | 36 | 39 | 44 | 47 | 44 | 47 | 60 | 63 | 60 | 63 | 60 | 63 | 60 | 63 | 36 | 39 | 36 | 39 | 44 | 47 | 44 | 47 | 60 | 63 | 60 | 63 | 60 | 63 | 60 | 63 | 37 | 39 | 37 | 39 | 45 | 47 | 45 | 47 | 61 | 63 | 61 | 63 | 61 | 63 | 61 | 63 | 37 | 39 | 37 | 39 | 45 | 47 | 45 | 47 | 61 | 63 | 61 | 63 | 61 | 63 | 61 | 63 |
| 37 | 39 | 37 | 39 | 45 | 47 | 45 | 47 | 61 | 63 | 61 | 63 | 61 | 63 | 61 | 63 | 37 | 39 | 37 | 39 | 45 | 47 | 45 | 47 | 61 | 63 | 61 | 63 | 61 | 63 | 61 | 63 | 37 | 39 | 37 | 39 | 45 | 47 | 45 | 47 | 61 | 63 | 61 | 63 | 61 | 63 | 61 | 63 | 37 | 39 | 37 | 39 | 45 | 47 | 45 | 47 | 61 | 63 | 61 | 63 | 61 | 63 | 61 | 63 |
| 38 | 38 | 38 | 38 | 46 | 46 | 46 | 46 | 62 | 62 | 62 | 62 | 62 | 62 | 62 | 62 | 38 | 38 | 38 | 38 | 46 | 46 | 46 | 46 | 62 | 62 | 62 | 62 | 62 | 62 | 62 | 62 | 39 | 39 | 39 | 39 | 47 | 47 | 47 | 47 | 63 | 63 | 63 | 63 | 63 | 63 | 63 | 63 | 39 | 39 | 39 | 39 | 47 | 47 | 47 | 47 | 63 | 63 | 63 | 63 | 63 | 63 | 63 | 63 |
| 39 | 39 | 39 | 39 | 47 | 47 | 47 | 47 | 63 | 63 | 63 | 63 | 63 | 63 | 63 | 63 | 39 | 39 | 39 | 39 | 47 | 47 | 47 | 47 | 63 | 63 | 63 | 63 | 63 | 63 | 63 | 63 | 39 | 39 | 39 | 39 | 47 | 47 | 47 | 47 | 63 | 63 | 63 | 63 | 63 | 63 | 63 | 63 | 39 | 39 | 39 | 39 | 47 | 47 | 47 | 47 | 63 | 63 | 63 | 63 | 63 | 63 | 63 | 63 |
| 40 | 43 | 46 | 46 | 40 | 43 | 46 | 46 | 56 | 59 | 62 | 62 | 56 | 59 | 62 | 62 | 40 | 43 | 46 | 46 | 40 | 43 | 46 | 46 | 56 | 59 | 62 | 62 | 56 | 59 | 62 | 62 | 41 | 43 | 47 | 47 | 41 | 43 | 47 | 47 | 57 | 59 | 63 | 63 | 57 | 59 | 63 | 63 | 41 | 43 | 47 | 47 | 41 | 43 | 47 | 47 | 57 | 59 | 63 | 63 | 57 | 59 | 63 | 63 |
| 41 | 43 | 47 | 47 | 41 | 43 | 47 | 47 | 57 | 59 | 63 | 63 | 57 | 59 | 63 | 63 | 41 | 43 | 47 | 47 | 41 | 43 | 47 | 47 | 57 | 59 | 63 | 63 | 57 | 59 | 63 | 63 | 41 | 43 | 47 | 47 | 41 | 43 | 47 | 47 | 57 | 59 | 63 | 63 | 57 | 59 | 63 | 63 | 41 | 43 | 47 | 47 | 41 | 43 | 47 | 47 | 57 | 59 | 63 | 63 | 57 | 59 | 63 | 63 |
| 42 | 42 | 46 | 46 | 42 | 42 | 46 | 46 | 58 | 58 | 62 | 62 | 58 | 58 | 62 | 62 | 42 | 42 | 46 | 46 | 42 | 42 | 46 | 46 | 58 | 58 | 62 | 62 | 58 | 58 | 62 | 62 | 43 | 43 | 47 | 47 | 43 | 43 | 47 | 47 | 59 | 59 | 63 | 63 | 59 | 59 | 63 | 63 | 43 | 43 | 47 | 47 | 43 | 43 | 47 | 47 | 59 | 59 | 63 | 63 | 59 | 59 | 63 | 63 |
| 43 | 43 | 47 | 47 | 43 | 43 | 47 | 47 | 59 | 59 | 63 | 63 | 59 | 59 | 63 | 63 | 43 | 43 | 47 | 47 | 43 | 43 | 47 | 47 | 59 | 59 | 63 | 63 | 59 | 59 | 63 | 63 | 43 | 43 | 47 | 47 | 43 | 43 | 47 | 47 | 59 | 59 | 63 | 63 | 59 | 59 | 63 | 63 | 43 | 43 | 47 | 47 | 43 | 43 | 47 | 47 | 59 | 59 | 63 | 63 | 59 | 59 | 63 | 63 |
| 44 | 47 | 44 | 47 | 44 | 47 | 44 | 47 | 60 | 63 | 60 | 63 | 60 | 63 | 60 | 63 | 44 | 47 | 44 | 47 | 44 | 47 | 44 | 47 | 60 | 63 | 60 | 63 | 60 | 63 | 60 | 63 | 45 | 47 | 45 | 47 | 45 | 47 | 45 | 47 | 61 | 63 | 61 | 63 | 61 | 63 | 61 | 63 | 45 | 47 | 45 | 47 | 45 | 47 | 45 | 47 | 61 | 63 | 61 | 63 | 61 | 63 | 61 | 63 |
| 45 | 47 | 45 | 47 | 45 | 47 | 45 | 47 | 61 | 63 | 61 | 63 | 61 | 63 | 61 | 63 | 45 | 47 | 45 | 47 | 45 | 47 | 45 | 47 | 61 | 63 | 61 | 63 | 61 | 63 | 61 | 63 | 45 | 47 | 45 | 47 | 45 | 47 | 45 | 47 | 61 | 63 | 61 | 63 | 61 | 63 | 61 | 63 | 45 | 47 | 45 | 47 | 45 | 47 | 45 | 47 | 61 | 63 | 61 | 63 | 61 | 63 | 61 | 63 |
| 46 | 46 | 46 | 46 | 46 | 46 | 46 | 46 | 62 | 62 | 62 | 62 | 62 | 62 | 62 | 62 | 46 | 46 | 46 | 46 | 46 | 46 | 46 | 46 | 62 | 62 | 62 | 62 | 62 | 62 | 62 | 62 | 47 | 47 | 47 | 47 | 47 | 47 | 47 | 47 | 63 | 63 | 63 | 63 | 63 | 63 | 63 | 63 | 47 | 47 | 47 | 47 | 47 | 47 | 47 | 47 | 63 | 63 | 63 | 63 | 63 | 63 | 63 | 63 |
| 47 | 47 | 47 | 47 | 47 | 47 | 47 | 47 | 63 | 63 | 63 | 63 | 63 | 63 | 63 | 63 | 47 | 47 | 47 | 47 | 47 | 47 | 47 | 47 | 63 | 63 | 63 | 63 | 63 | 63 | 63 | 63 | 47 | 47 | 47 | 47 | 47 | 47 | 47 | 47 | 63 | 63 | 63 | 63 | 63 | 63 | 63 | 63 | 47 | 47 | 47 | 47 | 47 | 47 | 47 | 47 | 63 | 63 | 63 | 63 | 63 | 63 | 63 | 63 |
| 48 | 51 | 54 | 54 | 60 | 63 | 60 | 63 | 48 | 51 | 54 | 54 | 60 | 63 | 60 | 63 | 48 | 51 | 54 | 54 | 60 | 63 | 60 | 63 | 48 | 51 | 54 | 54 | 60 | 63 | 60 | 63 | 49 | 51 | 55 | 55 | 61 | 63 | 61 | 63 | 49 | 51 | 55 | 55 | 61 | 63 | 61 | 63 | 49 | 51 | 55 | 55 | 61 | 63 | 61 | 63 | 49 | 51 | 55 | 55 | 61 | 63 | 61 | 63 |
| 49 | 51 | 55 | 55 | 61 | 63 | 61 | 63 | 49 | 51 | 55 | 55 | 61 | 63 | 61 | 63 | 49 | 51 | 55 | 55 | 61 | 63 | 61 | 63 | 49 | 51 | 55 | 55 | 61 | 63 | 61 | 63 | 49 | 51 | 55 | 55 | 61 | 63 | 61 | 63 | 49 | 51 | 55 | 55 | 61 | 63 | 61 | 63 | 49 | 51 | 55 | 55 | 61 | 63 | 61 | 63 | 49 | 51 | 55 | 55 | 61 | 63 | 61 | 63 |
| 50 | 50 | 54 | 54 | 62 | 62 | 62 | 62 | 50 | 50 | 54 | 54 | 62 | 62 | 62 | 62 | 50 | 50 | 54 | 54 | 62 | 62 | 62 | 62 | 50 | 50 | 54 | 54 | 62 | 62 | 62 | 62 | 51 | 51 | 55 | 55 | 63 | 63 | 63 | 63 | 51 | 51 | 55 | 55 | 63 | 63 | 63 | 63 | 51 | 51 | 55 | 55 | 63 | 63 | 63 | 63 | 51 | 51 | 55 | 55 | 63 | 63 | 63 | 63 |
| 51 | 51 | 55 | 55 | 63 | 63 | 63 | 63 | 51 | 51 | 55 | 55 | 63 | 63 | 63 | 63 | 51 | 51 | 55 | 55 | 63 | 63 | 63 | 63 | 51 | 51 | 55 | 55 | 63 | 63 | 63 | 63 | 51 | 51 | 55 | 55 | 63 | 63 | 63 | 63 | 51 | 51 | 55 | 55 | 63 | 63 | 63 | 63 | 51 | 51 | 55 | 55 | 63 | 63 | 63 | 63 | 51 | 51 | 55 | 55 | 63 | 63 | 63 | 63 |
| 52 | 55 | 52 | 55 | 60 | 63 | 60 | 63 | 52 | 55 | 52 | 55 | 60 | 63 | 60 | 63 | 52 | 55 | 52 | 55 | 60 | 63 | 60 | 63 | 52 | 55 | 52 | 55 | 60 | 63 | 60 | 63 | 53 | 55 | 53 | 55 | 61 | 63 | 61 | 63 | 53 | 55 | 53 | 55 | 61 | 63 | 61 | 63 | 53 | 55 | 53 | 55 | 61 | 63 | 61 | 63 | 53 | 55 | 53 | 55 | 61 | 63 | 61 | 63 |
| 53 | 55 | 53 | 55 | 61 | 63 | 61 | 63 | 53 | 55 | 53 | 55 | 61 | 63 | 61 | 63 | 53 | 55 | 53 | 55 | 61 | 63 | 61 | 63 | 53 | 55 | 53 | 55 | 61 | 63 | 61 | 63 | 53 | 55 | 53 | 55 | 61 | 63 | 61 | 63 | 53 | 55 | 53 | 55 | 61 | 63 | 61 | 63 | 53 | 55 | 53 | 55 | 61 | 63 | 61 | 63 | 53 | 55 | 53 | 55 | 61 | 63 | 61 | 63 |
| 54 | 54 | 54 | 54 | 62 | 62 | 62 | 62 | 54 | 54 | 54 | 54 | 62 | 62 | 62 | 62 | 54 | 54 | 54 | 54 | 62 | 62 | 62 | 62 | 54 | 54 | 54 | 54 | 62 | 62 | 62 | 62 | 55 | 55 | 55 | 55 | 63 | 63 | 63 | 63 | 55 | 55 | 55 | 55 | 63 | 63 | 63 | 63 | 55 | 55 | 55 | 55 | 63 | 63 | 63 | 63 | 55 | 55 | 55 | 55 | 63 | 63 | 63 | 63 |
| 55 | 55 | 55 | 55 | 63 | 63 | 63 | 63 | 55 | 55 | 55 | 55 | 63 | 63 | 63 | 63 | 55 | 55 | 55 | 55 | 63 | 63 | 63 | 63 | 55 | 55 | 55 | 55 | 63 | 63 | 63 | 63 | 55 | 55 | 55 | 55 | 63 | 63 | 63 | 63 | 55 | 55 | 55 | 55 | 63 | 63 | 63 | 63 | 55 | 55 | 55 | 55 | 63 | 63 | 63 | 63 | 55 | 55 | 55 | 55 | 63 | 63 | 63 | 63 |
| 56 | 59 | 62 | 62 | 56 | 59 | 62 | 62 | 56 | 59 | 62 | 62 | 56 | 59 | 62 | 62 | 56 | 59 | 62 | 62 | 56 | 59 | 62 | 62 | 56 | 59 | 62 | 62 | 56 | 59 | 62 | 62 | 57 | 59 | 63 | 63 | 57 | 59 | 63 | 63 | 57 | 59 | 63 | 63 | 57 | 59 | 63 | 63 | 57 | 59 | 63 | 63 | 57 | 59 | 63 | 63 | 57 | 59 | 63 | 63 | 57 | 59 | 63 | 63 |
| 57 | 59 | 63 | 63 | 57 | 59 | 63 | 63 | 57 | 59 | 63 | 63 | 57 | 59 | 63 | 63 | 57 | 59 | 63 | 63 | 57 | 59 | 63 | 63 | 57 | 59 | 63 | 63 | 57 | 59 | 63 | 63 | 57 | 59 | 63 | 63 | 57 | 59 | 63 | 63 | 57 | 59 | 63 | 63 | 57 | 59 | 63 | 63 | 57 | 59 | 63 | 63 | 57 | 59 | 63 | 63 | 57 | 59 | 63 | 63 | 57 | 59 | 63 | 63 |
| 58 | 58 | 62 | 62 | 58 | 58 | 62 | 62 | 58 | 58 | 62 | 62 | 58 | 58 | 62 | 62 | 58 | 58 | 62 | 62 | 58 | 58 | 62 | 62 | 58 | 58 | 62 | 62 | 58 | 58 | 62 | 62 | 59 | 59 | 63 | 63 | 59 | 59 | 63 | 63 | 59 | 59 | 63 | 63 | 59 | 59 | 63 | 63 | 59 | 59 | 63 | 63 | 59 | 59 | 63 | 63 | 59 | 59 | 63 | 63 | 59 | 59 | 63 | 63 |
| 59 | 59 | 63 | 63 | 59 | 59 | 63 | 63 | 59 | 59 | 63 | 63 | 59 | 59 | 63 | 63 | 59 | 59 | 63 | 63 | 59 | 59 | 63 | 63 | 59 | 59 | 63 | 63 | 59 | 59 | 63 | 63 | 59 | 59 | 63 | 63 | 59 | 59 | 63 | 63 | 59 | 59 | 63 | 63 | 59 | 59 | 63 | 63 | 59 | 59 | 63 | 63 | 59 | 59 | 63 | 63 | 59 | 59 | 63 | 63 | 59 | 59 | 63 | 63 |
| 60 | 63 | 60 | 63 | 60 | 63 | 60 | 63 | 60 | 63 | 60 | 63 | 60 | 63 | 60 | 63 | 60 | 63 | 60 | 63 | 60 | 63 | 60 | 63 | 60 | 63 | 60 | 63 | 60 | 63 | 60 | 63 | 61 | 63 | 61 | 63 | 61 | 63 | 61 | 63 | 61 | 63 | 61 | 63 | 61 | 63 | 61 | 63 | 61 | 63 | 61 | 63 | 61 | 63 | 61 | 63 | 61 | 63 | 61 | 63 | 61 | 63 | 61 | 63 |
| 61 | 63 | 61 | 63 | 61 | 63 | 61 | 63 | 61 | 63 | 61 | 63 | 61 | 63 | 61 | 63 | 61 | 63 | 61 | 63 | 61 | 63 | 61 | 63 | 61 | 63 | 61 | 63 | 61 | 63 | 61 | 63 | 61 | 63 | 61 | 63 | 61 | 63 | 61 | 63 | 61 | 63 | 61 | 63 | 61 | 63 | 61 | 63 | 61 | 63 | 61 | 63 | 61 | 63 | 61 | 63 | 61 | 63 | 61 | 63 | 61 | 63 | 61 | 63 |
| 62 | 62 | 62 | 62 | 62 | 62 | 62 | 62 | 62 | 62 | 62 | 62 | 62 | 62 | 62 | 62 | 62 | 62 | 62 | 62 | 62 | 62 | 62 | 62 | 62 | 62 | 62 | 62 | 62 | 62 | 62 | 62 | 63 | 63 | 63 | 63 | 63 | 63 | 63 | 63 | 63 | 63 | 63 | 63 | 63 | 63 | 63 | 63 | 63 | 63 | 63 | 63 | 63 | 63 | 63 | 63 | 63 | 63 | 63 | 63 | 63 | 63 | 63 | 63 |
| 63 | 63 | 63 | 63 | 63 | 63 | 63 | 63 | 63 | 63 | 63 | 63 | 63 | 63 | 63 | 63 | 63 | 63 | 63 | 63 | 63 | 63 | 63 | 63 | 63 | 63 | 63 | 63 | 63 | 63 | 63 | 63 | 63 | 63 | 63 | 63 | 63 | 63 | 63 | 63 | 63 | 63 | 63 | 63 | 63 | 63 | 63 | 63 | 63 | 63 | 63 | 63 | 63 | 63 | 63 | 63 | 63 | 63 | 63 | 63 | 63 | 63 | 63 | 63 |

Figure 11: Matrix $M_{L_4}$ of composition $\mathbf{x} \circ \mathbf{y}$ implemented by $L_4$. Symmetric entries are underlined.